\documentclass[twocolumn]{revtex4}
\usepackage[utf8]{inputenc}
\usepackage{graphicx}

\begin{document}

\title{On the generalized Keffer form of the Dzyaloshinskii constant:
its consequences for the spin, momentum and polarization evolution}

\author{Pavel A. Andreev}
\email{andreevpa@my.msu.ru}
\affiliation{Department of General Physics, Faculty of physics, Lomonosov Moscow State University, Moscow, Russian Federation, 119991.}


\begin{abstract}
Different analytical features of the Dzyaloshinskii-Moriya interaction are related to different contributions to the Dzyaloshinskii constant in the microscopic Hamiltonian.
Consequences appear in the macroscopic Landau--Lifshitz--Gilbert equation.
It leads to various phenomena.
Three contributions to the Dzyaloshinskii constant are reviewed
and combined in the generalized Keffer form of the Dzyaloshinskii constant.
The fourth possible form of the Dzyaloshinskii constant is suggested as well.
Macroscopic consequences of these three mechanisms are well-known,
but further possible generalizations of the Keffer form of the Dzyaloshinskii constant are suggested.
Consequences for the spin evolution equations,
the momentum balance equations,
and polarization evolution equations are considered.
Some analog of the Keffer form is suggested for the exchange integral in symmetric Heisenberg Hamiltonian demonstrating the nontrivial contribution of the ligands in this regime.
\end{abstract}


\maketitle



\section{Introduction}

The Dzyaloshinskii-Moriya interaction (DMI) plays important role in the understanding of the nontrivial magnetic structures in condensed matter physics.
Basically, the DMI can be presented within the Hamiltonian for the effective interaction of the magnetic moments
$\hat{\mbox{\boldmath $\mu$}}$ in the medium
\cite{Dzyaloshinskii JETP 57}, \cite{Dzyaloshinskii JETP 64}, \cite{Moriya PR 60}, \cite{Fert PRL 80}
$\hat{H}_{DMI}=\textbf{D}_{12}\cdot[\hat{\textbf{S}}_{1}\times \hat{\textbf{S}}_{2}]$,
where the spin operators $\hat{\textbf{S}}_{i}$ are applied,
$\hat{\mbox{\boldmath $\mu$}}_{i}\sim\hat{\textbf{S}}_{i}$,
$\textbf{D}_{12}=-\textbf{D}_{21}$ is the Dzyaloshinskii-Moriya constant (DMC),
$i=1,2$ are the numbers of particles.
The DMI has nature of the exchange interaction similarly to the Heisenberg exchange interaction,
but its existence requires the break of some symmetries in the system.
The exchange interaction between magnetic ions in oxides and other structures composed of several types of ions happens nondirectly.
There is the exchange interaction of magnetic ions with the nonmagnetic ions
since the nonmagnetic ions are located between the magnetic ions.
This mechanism is also called the superexchange interaction.
In contrast to the simple isotropic exchange interaction
(or the symmetric Heisenberg interaction)
$\hat{H}_{SHI}=U_{12}(r_{12})(\hat{\textbf{S}}_{1}\cdot \hat{\textbf{S}}_{2})$,
where
$r_{12}$ is the distance between the magnetic moments/spins,
the superexchange interaction (SEI) shows the anisotropy
and scalar function $U_{12}(r_{12})$ becomes a matrix
$\hat{U}_{12}(\textbf{r}_{12})$
(the second rank matrix bonding the spin vectors)
$\hat{H}_{SEI}=(\hat{\textbf{S}}_{1}\cdot \hat{U}_{12}(\textbf{r}_{12})\cdot\hat{\textbf{S}}_{2})$
$=U_{12}^{\alpha\beta}(\textbf{r}_{12})\hat{S}_{1}^{\alpha}\hat{S}_{2}^{\beta}$,
where the Greek indexes like $\alpha=x,y,z$ correspond to projection to the coordinate axis,
the summation on the repeating indexes is assumed.
Here, matrix $\hat{U}_{12}(\textbf{r}_{12})$ is an arbitrary nonsymmetric matrix
depending on the relative position of the magnetic ions $\textbf{r}_{12}$ instead of the distance described by the module of this vector $r_{12}=\mid\textbf{r}_{12}\mid$.
This matrix $\hat{U}_{12}(\textbf{r}_{12})$ can be represented as the sum of its symmetric and antisymmetric parts:
$U_{12}^{\alpha\beta}(\textbf{r}_{12})=U_{12,S}^{\alpha\beta}(\textbf{r}_{12})+U_{12,A}^{\alpha\beta}(\textbf{r}_{12})$,
where
$U_{12,S}^{\alpha\beta}(\textbf{r}_{12})=(U_{12}^{\alpha\beta}(\textbf{r}_{12})+U_{12}^{\beta\alpha}(\textbf{r}_{12}))/2$ is the symmetric part,
and
$U_{12,A}^{\alpha\beta}(\textbf{r}_{12})=(U_{12}^{\alpha\beta}(\textbf{r}_{12})-U_{12}^{\beta\alpha}(\textbf{r}_{12}))/2$ is the antisymmetric part.
The symmetric part can be diagonalized by choosing corresponding axis $\hat{U}_{12,S}(\textbf{r}_{12})=\{U_{xx},U_{yy},U_{zz}\}$.
It gives the $X,Y,Z$-model.
The antisymmetric part contains three nonzero values
$$\hat{U}_{12,A}(\textbf{r}_{12})=\left(
                                    \begin{array}{ccc}
                                      0 & U_{12,A} & -U_{13,A} \\
                                      -U_{12,A} & 0 & U_{23,A} \\
                                      U_{13,A} & -U_{23,A} & 0 \\
                                    \end{array}
                                  \right).
$$
Therefore, it can be represented as the "vector" (pseudovector or axial vector) $\textbf{D}_{12}$:
$U_{12,A}^{\alpha\beta}(\textbf{r}_{12})=\varepsilon^{\alpha\beta\gamma}D^{\gamma}_{12}(\textbf{r}_{12})$,
or $D^{\alpha}_{12}(\textbf{r}_{12})=(1/2)\varepsilon^{\alpha\beta\gamma}U_{12,A}^{\beta\gamma}(\textbf{r}_{12})$,
where
$\varepsilon^{\alpha\beta\gamma}$ is the unit antisymmetric third rank tensor
(the Levi-Civita symbol).
Hence, the second part represents the Dzyaloshinskii-Moriya interaction.

It happens that the DMC structure in the microscopic Hamiltonian highly influences the macroscopic values describing DMI in the limit of continuous medium
(such as the energy density, the spin torques, force density in the Euler equation for the momentum density, etc).
Therefore, it is necessary to present systematic analysis of the possible structures of the DMC.
Most famous expression for the DMC is the Keffer form
\cite{Fishman PRB 19}, \cite{Khomskii JETP 21}
$$\textbf{D}_{ij}=
\beta(r_{ij})[\textbf{r}_{ij}\times\mbox{\boldmath $\delta$}_{1}],$$
where the antisymmetry of the DMC formed by the antisymmetry of relative distance of the magnetic ions $\textbf{r}_{ij}$,
and the ligand shift from the position of the center of mass of charges is also applied $\mbox{\boldmath $\delta$}_{1}$.
Different representation of the structure of the Keffer form of DMC can be found in Ref. \cite{Pyatakov UFN 12}.
The direction of DMC $\textbf{D}_{ij}$ is perpendicular to the plane,
where two magnetic ions and the ligand are located (like \cite{Moon PRB 13} eq. 1 and text after eq. 1).
However, direct consequences of this DMC are less known in relation to the macroscopic DMI.
Similar equations are usually considered as a part of the Lifshitz invariant
(see Ref. \cite{Gareeva PRB 13} eq. 5, and \cite{Sparavigna PRB 94} eq. 8).
Derivation of the energy density related to the Keffer can be found in Ref.
\cite{ZvezdinPyatakov EPL 12}.
Further application of the Keffer form for DMC to the spin torque and the spin current model can be found
in Refs. \cite{AndreevTrukh JETP 24} and \cite{AndreevTrukh PS 24}.
Macroscopic energy density can be also found in Refs.
\cite{Camley SSR 23} (see eq. 6) and \cite{Gobel PR 21},
where the contribution of the ligand shift is included in unexplicit way.
As a result they have a scalar macroscopic DMC and all spin densities are given via projections,
so, the vector or tensor structure is hidden.
In fig 2b of Ref. \cite{Camley SSR 23}, authors show the direction of DMC, which allows to complete the structure of eq. 6 of their paper.

Another form for the DMC is
$$\textbf{D}_{ij}=
\gamma(r_{ij})\textbf{r}_{ij}$$
(see \cite{Fishman PRB 19}).
It also leads to a part of Lifshitz invariant in the macroscopic limit.
This form does not contain any explicit trace of the ligand ion,
while ligands may be necessary for the formation of the required symmetry conditions in the medium.
Its macroscopic form is well-known, so the free energy can be found in textbooks like \cite{Landau 8}
(see Sec. 52).
It can be also found in recent applications (see Ref. \cite{Castillo-Sepulveda RiP 24}).

Effective interaction proportional to the vector product of the partial spin densities in the two-component antiferromagnetic materials as a part of the free-energy $\sim [\textbf{S}_{A}\times\textbf{S}_{B}]$
was suggested
by Dzyaloshinskii \cite{Dzyaloshinskii JETP 57}
in order to describe
the weak ferromagnetizm
in the antiferromagnetic materials
(it can be also found in Ref. \cite{Landau 8} Sec. 50 eq. 1).
To complete the scalar expression for the energy density one requires unit vector $\textbf{n}$:
$\mathcal{E}=D \textbf{n}\cdot[\textbf{S}_{A}\times\textbf{S}_{B}]$,
where $D$ is the scalar part of the DMC $\textbf{D}=D \textbf{n}$.
This term is suggested for the magnetic materials with two magnetic subspecies
(like the antiferromagnetic materials).
So, it has no analogue for the ferromagnetic materials.
Some recent application of this term to the antiferromagnetic multiferroic materials can be found in Ref. \cite{Gareeva PRB 13} (eq. 2 and 3).
Particularly, phase diagrams are analyzed in Ref. \cite{Gareeva PRB 13}.

Dzyaloshinskii suggested that the equilibrium uncompensated magnetization of the antiferromagnetic medium  is related to the reverse magneto-electric effect.
It happens due to the electric electric dipole moment,
which changes the direction at the transition from one crystal cell to the next one.
The direction of the approximately "compensated" electric dipoles is used
in order to argue the existence of the vector constant $\textbf{n}$ in the free energy.
So, the appearance of the uncompensated magnetization is related to
the existence of electric quadrupoles.

Existence of DMI is important for the magnetoelectric effect.
Particularly, the DMI tends to reorganize the magnetic moments in noncollinear structures,
while the noncollinear order of spins
via the force acting on ions due to combined action of the spin-orbit and symmetric Heisenberg exchange interactions
leads to the formation of polarization, as one of possible mechanisms \cite{AndreevTrukh PS 24}.
Another mechanism involves the DMI in a different way.
No matter collinear or partially noncollinear structure of spins is formed in the medium,
the combination of the spin-orbit and Dzyaloshinskii-Moriya interactions gives the force field acting on ions and forming the polarization related to the collinear parts of spins \cite{AndreevTrukh PS 24}.
Describing the polarization in this paper we consider the polarization of spin origin, like described here.
Hence, magneto-electric effects have different origins.
There are also the current-driven mechanisms \cite{Tatara PR 08} along with the DMI presented above.

One of remarkable dynamical manifestations of the magneto-electric effects is
the electromagnon resonance discovered experimentally \cite{Pimenov NP 06}, \cite{ShuvaevPimenov EPJB 11}.
Its theoretical analysis and connection between the magneto-electric effect and the electromagnon resonance is suggested in Ref. \cite{Katsura PRL 07}.
Recent application of this concept to in the non-cycloidal phase of TbMnO$_{3}$ can be found in Ref. \cite{Aupiais npj QM 18}.
Some additional discussions can be found in review article \cite{Tokura RPP 14}.
Resent analytical approach to the main mechanisms of the electromagnon resonance can be also found in Ref. \cite{Andreev 2025 06}.

This paper is organized as follows.
In Sec. II the possible structures of DMC is discussed.
In Sec. III the contributions of different forms of the DMC in the spin torques existing in the Landau--Lifshitz--Gilbert equation is demonstrated.
In Sec. IV the energy densities are obtained and compared for different parts of the DMI described by different DMC.
In Sec. V the magnetoelectric effect is discussed,
some forms of the polarization of medium are described
in relation to the spin-current model and physical mechanisms (interactions) leading to the polarization formation.
Corresponding spin torques are shown as well.
In Sec. VI the contribution of the DMI in the polarization evolution equation is demonstrated.
In Sec. VII the force density in the Euler equation for the momentum density is derived for different parts of the DMI.
In Sec. VIII generalized symmetric Heisenberg interaction appearing at the account of the ligand shift is suggested.
In Sec. IX a brief summary of obtained results is presented.

\begin{figure}\includegraphics[width=8cm,angle=0]{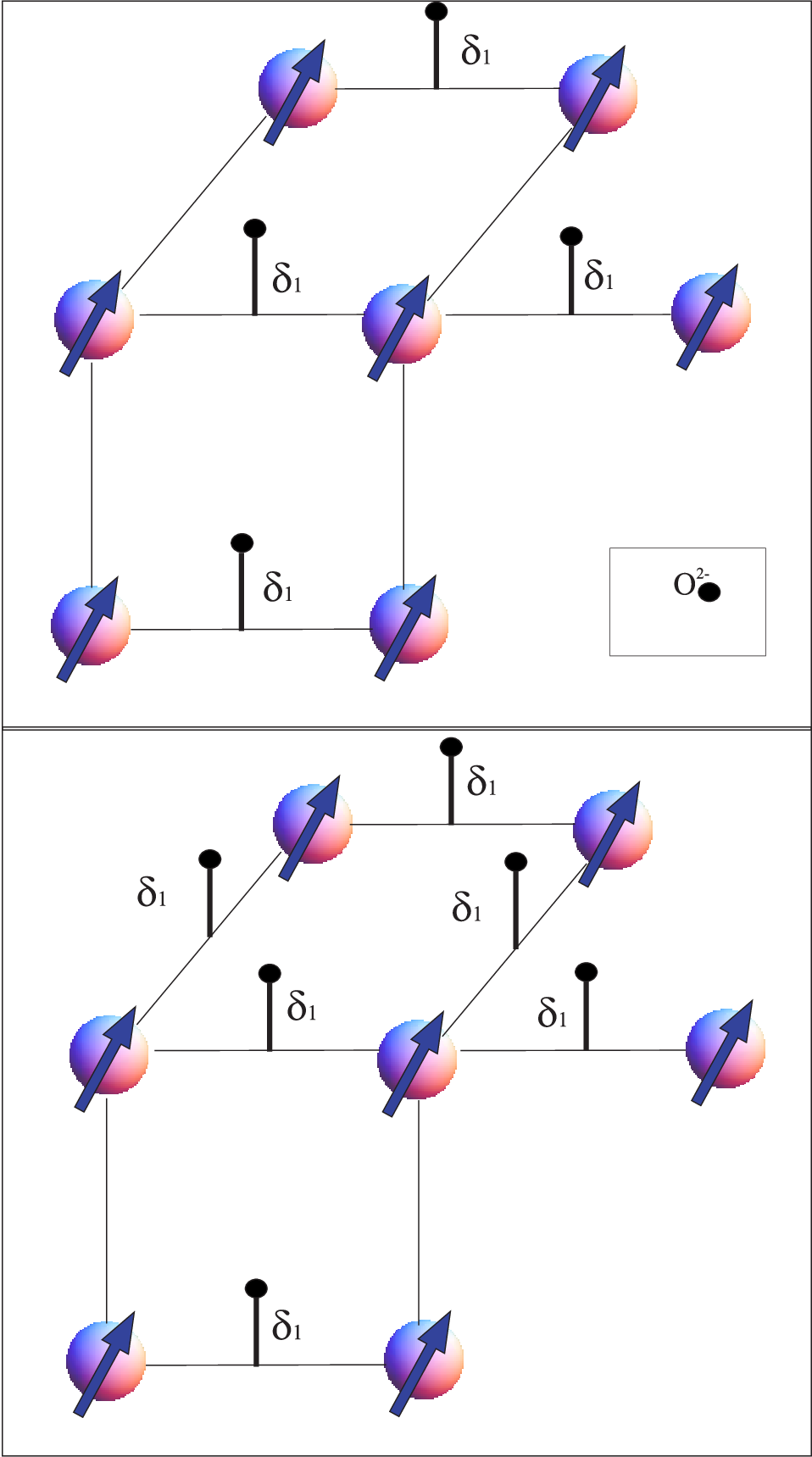}
\caption{\label{MFMextremumP Fig 01} The figure shows
two possibilities for the location of the ligand ions relatively magnetic ions
and the possibilities of their shifts from the positions corresponding to the middle point between magnetic ions.
Direction of the magnetic moments is fixed along "vertical" direction,
but this is unnecessary choice,
the anisotropy axis is not presented in the figure,
type of the magnetic (easy axis or easy plane) is not fixed either.
Same picture can be made for the antiferromagnetic materials.}
\end{figure}

\begin{figure}\includegraphics[width=8cm,angle=0]{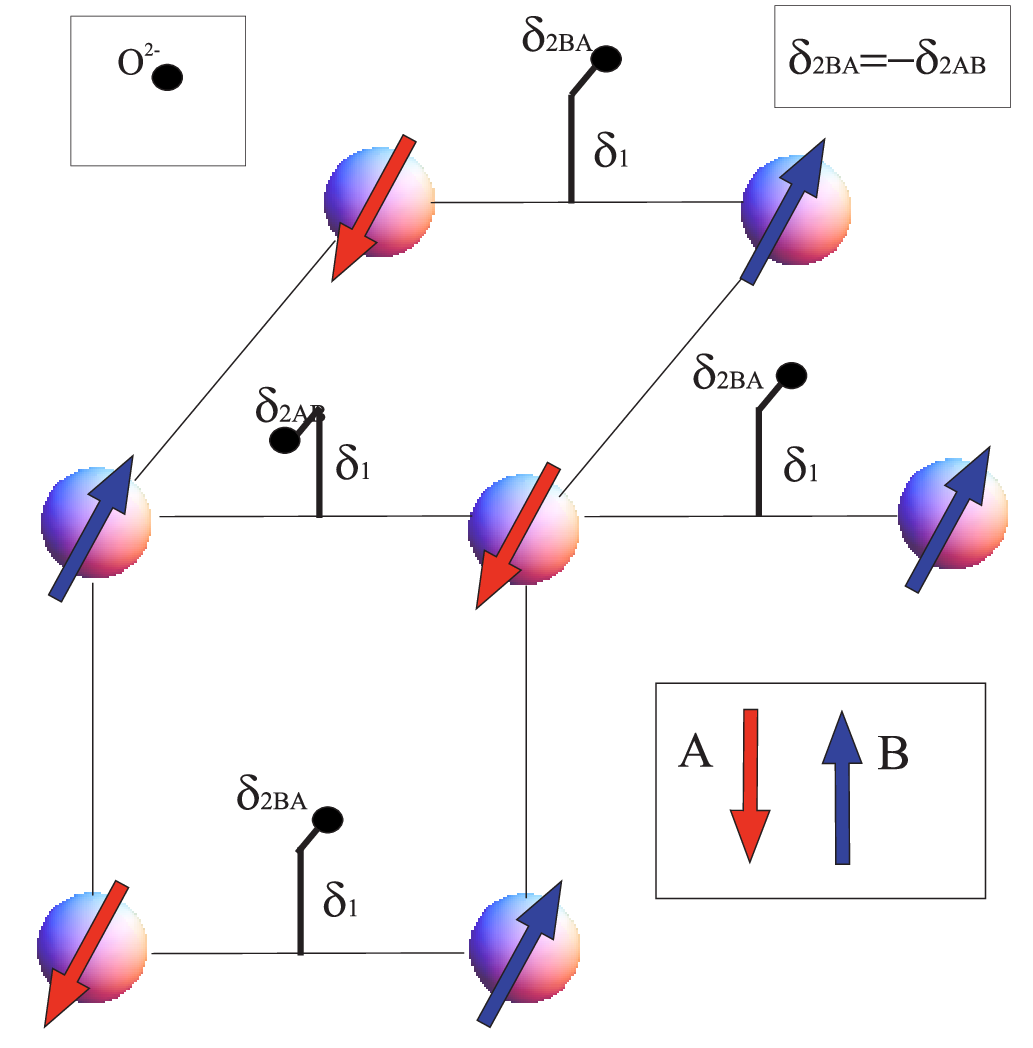}
\caption{\label{MFMextremumP Fig 02} The figure shows
a possibility for the location of the ligand ions relatively magnetic ions
in the antiferromagnetic materials, more complex in comparison with Fig. (\ref{MFMextremumP Fig 01}).}
\end{figure}

\section{Dzyaloshinskii-Moriya interaction and the Keffer form of the Dzyaloshinskii constant}

The Hamiltonian of the Dzyaloshinskii-Moriya interaction can be written in the well-known form
\begin{equation}\label{MFMemf Ham DMI}
\hat{H}=(-1/2)\sum_{i,j,j\neq i}(\textbf{D}_{ij}\cdot[\hat{\textbf{S}}_{i}\times \hat{\textbf{S}}_{j}]),
\end{equation}
where $i$ and $j$ are the numbers of particles,
$\textbf{D}_{ij}$ is the Dzyaloshinskii constant,
$\hat{\textbf{S}}_{i}$ is the vector operator of spin of $i$-th particle,
symbol $[\textbf{a}\times \textbf{b}]$ is the vector product of vectors $\textbf{a}$ and $\textbf{b}$,
symbol $(\textbf{a}\cdot \textbf{b})$ is the scalar product of vectors $\textbf{a}$ and $\textbf{b}$.
Let us present the commutation relations for the projections of spin operators
$[\hat{S}_{i}^{\alpha},\hat{S}_{i}^{\beta}]=\imath\hbar\varepsilon^{\alpha\beta\gamma}\hat{S}_{i}^{\gamma}$,
where it is given in the tensor notations.
Hence,
$\alpha$ and other Greek subindexes are the tensor indexes equal to $x,y,z$,
$\hat{S}_{i}^{\alpha}$ is the vector operator of spin of $i$-th particle corresponding to $\alpha$ projection,
$\imath$ is the imaginary unit $\imath^{2}=-1$,
$\hbar$ is the Planck constant,
$\varepsilon^{\alpha\beta\gamma}$ is the unit antisymmetric third rank tensor
(Levi-Civita symbol),
and the summation on the repeating indexes is assumed.

\subsection{Keffer form of the Dzyaloshinskii constant}

Let us discuss the analytical form of the Dzyaloshinskii constant.
In the simplest case it can be considered as the relative distance between magnetic ions
$\textbf{D}_{ij}=\gamma(r_{ij})\textbf{r}_{ij}$.
It leads to the energy density and corresponding spin torque describing
helicoidal structures \cite{Landau 8}.

Another form of the Dzyaloshinskii constant is suggested by Keffer
\cite{Keffer PR 62}, \cite{Moskvin PSS 77}.
This form includes the shift of the ligand from the position of center of mass/charge in the two neighboring magnetic ions:
$\textbf{D}_{ij}=\beta(r_{ij})[\textbf{r}_{ij}\times\mbox{\boldmath $\delta$}_{1}]$
(see also \cite{Fishman PRB 19} and \cite{Khomskii JETP 21}).
The ligand shift is illustrated in Fig. \ref{MFMextremumP Fig 01}.

Both mechanisms can be combined in the single equation
\cite{Fishman PRB 19}
\begin{equation}\label{MFMemf D Dz const str 0}
\textbf{D}_{ij}=\gamma(r_{ij})\textbf{r}_{ij}
+\beta(r_{ij})[\textbf{r}_{ij}\times\mbox{\boldmath $\delta$}_{1}]. \end{equation}
Corresponding spin torque is applied for the analysis of the spin wave dispersion dependence in the antiferromagnetic materials
if the equilibrium is described by the cycloidal order of spins
\cite{Andreev 25 10}.
Particularly, it is mentioned in \cite{Andreev 25 10} that the second term in (\ref{MFMemf D Dz const str 0}) is in agreement with the cycloidal order,
while the first term can correspond to the helicoidal order.
Obviously, the torque corresponding to different structures of the Dzyaloshinskii constant are different
and the energy densities are different as well.
However, non of these forms allows to obtain the energy density originally suggested by
Dzyaloshinskii
(it can be also found
\cite{Landau 8}).
Hence, equation (\ref{MFMemf D Dz const str 0}) should be generalized.
Let us to point out that
equation (\ref{MFMemf D Dz const str 0}) can be applied both to the ferromagnetic and antiferromagnetic materials.
Schematic illustration for possible positions of the magnetic ions and ligands is presented in Fig. \ref{MFMextremumP Fig 01}.
Moreover, if the ligand shift can affect the constant of the exchange interaction
(the Dzyaloshinskii-Moriya interaction is basically the "antisymmetric" part of the superexchange),
it can give contribution in the form of coefficient in the the "symmetric" part of the exchange interaction
(in the symmetric Heisenberg Hamiltonian, in other words).
This concept is considered below in the
final section
of this paper.

\subsection{Generalization of the Keffer form of the Dzyaloshinskii constant}

Here we suggest the Dzyaloshinskii constant composed of four group of terms
$$\textbf{D}_{ij}=\gamma(r_{ij})\textbf{r}_{ij}
+\beta(r_{ij})[\textbf{r}_{ij}\times\mbox{\boldmath $\delta$}_{1}]
+\tilde{\zeta}_{1}(r_{ij}) \mbox{\boldmath $\delta$}_{ij-AB}$$
\begin{equation}\label{MFMemf D Dz const str G}
+\tilde{\zeta}_{2}(r_{ij}) [[\textbf{r}_{ij}\times\mbox{\boldmath $\delta$}_{ij-AB}]\times\mbox{\boldmath $\delta$}_{ij-AB}],
\end{equation}
first two terms repeat equation (\ref{MFMemf D Dz const str 0}).
The third term is proportional to the ligand shift.
This term (as the full Dzyaloshinskii constant) should be an odd function relatively numbers $i$ and $j$.
Hence, it is possible for the antiferromagnetic materials,
where for the part of the ligand shift:
$\mbox{\boldmath $\delta$}_{2,ij-AB}=-\mbox{\boldmath $\delta$}_{2,ji-BA}$ (see Fig. \ref{MFMextremumP Fig 02})
or $\mbox{\boldmath $\delta$}_{3,ij-AB}=\mbox{\boldmath $\delta$}_{3,ji-BA}$ (see Fig. \ref{MFMextremumP Fig 03})
or full ligand shift $\mbox{\boldmath $\delta$}_{1,ij-AB}=-\mbox{\boldmath $\delta$}_{1,ji-BA}$ illustrated in Fig. \ref{MFMextremumP Fig 04}.
The last term is proportional to the double vector product and contains two ligand shifts.
It can be considered for both the ferromagnetic and antiferromagnetic materials,
but it gives nontrivial results at the account of two different shifts in the antiferromagnetic materials only.

The Keffer form of the Dzyaloshinskii constant can include the complete or partial shift $\mbox{\boldmath $\delta$}_{1}$
(it is complete shift for the regimes illustrated in Fig. \ref{MFMextremumP Fig 01},
but it is a partial shift for the regimes illustrated in Figs. \ref{MFMextremumP Fig 02} and \ref{MFMextremumP Fig 03}).
It is related to the fact that combinations like $[\textbf{r}_{ij}\times\mbox{\boldmath $\delta$}_{2,ij-AB}]$
are symmetric on the permutation $i\leftrightarrow j$.
Combinations like
$\beta'(r_{ij})[\mbox{\boldmath $\delta$}_{2,ij-AB}\times\mbox{\boldmath $\delta$}_{1}]$ (which is parallel to $\mbox{\boldmath $\delta$}_{3,ij-AB}$)
and
$\beta''(r_{ij})[\mbox{\boldmath $\delta$}_{3,ij-AB}\times\mbox{\boldmath $\delta$}_{1}]$ (which is parallel to $\mbox{\boldmath $\delta$}_{2,ij-AB}$)
are included in the third term in equation (\ref{MFMemf D Dz const str G})
which is discussed below in this section.

Our target is $\zeta_{1}(r_{ij})\mbox{\boldmath $\delta$}_{2,ij-AB}$.
"Oscillating" $\mbox{\boldmath $\delta$}_{2}$ gives required antisymmetry of the coefficients.
It gives the required term suggested by
Dzyaloshinskii.
This behavior of the ligand shift (the "oscillations") are possible for the antiferromagnetic materials only
(meaning it is impossible for the ferromagnetic materials,
while ferrimagnetic materials are similar to the antiferromagnetic materials and can support such ligand formation).

\subsection{The Dzyaloshinskii constant parallel to the ligand shifts}

After equation (\ref{MFMemf D Dz const str G})
we introduced three partial "oscillating" shifts of the ligands:
$\mbox{\boldmath $\delta$}_{2,ij-AB}=-\mbox{\boldmath $\delta$}_{2,ji-BA}$ (see Fig. \ref{MFMextremumP Fig 02})
or $\mbox{\boldmath $\delta$}_{3,ij-AB}=\mbox{\boldmath $\delta$}_{3,ji-BA}$ (see Fig. \ref{MFMextremumP Fig 03})
or $\mbox{\boldmath $\delta$}_{1,ij-AB}=-\mbox{\boldmath $\delta$}_{1,ji-BA}$ (see Fig. \ref{MFMextremumP Fig 04}).
We can also assume that pairs of "oscillating" shifts
can exist simultaneously
$\mbox{\boldmath $\delta$}_{1,ij-AB}$ and $\mbox{\boldmath $\delta$}_{2,ij-AB}$
or $\mbox{\boldmath $\delta$}_{1,ij-AB}$ and $\mbox{\boldmath $\delta$}_{3,ij-AB}$.
However, even one of them allows to consider the third term in the Dzyaloshinskii constant (\ref{MFMemf D Dz const str G}).
This part of the
the Dzyaloshinskii constant is parallel to the partial ligand shifts:
$$\textbf{D}_{ij,(\delta)}=\tilde{\zeta}_{1}(r_{ij})\mbox{\boldmath $\delta$}_{ij}=$$
\begin{equation}\label{MFMemf D Dz const str G par delta}
=\zeta_{1}(r_{ij})\mbox{\boldmath $\delta$}_{2,ij-AB}
+\zeta_{1}'(r_{ij})\mbox{\boldmath $\delta$}_{3,ij-AB}
+\zeta_{1}''(r_{ij})\mbox{\boldmath $\delta$}_{1,ij-AB}. \end{equation}
The part proportional to $\mbox{\boldmath $\delta$}_{2,ij-AB}$ presents main interest in historical perspective.
It should be considered in combination with fixed $\mbox{\boldmath $\delta$}_{1}$ as it is presented in Fig. \ref{MFMextremumP Fig 02}.
It reproduces the relative position of magnetic and nonmagnetic ions considered by
Dzyaloshinskii.
Further calculations and macroscopic equations have same structure for each of three partial terms in (\ref{MFMemf D Dz const str G par delta}).
Corresponding equations are derived below to show that presented form of Dzyaloshinskii constant reproduces well-known results
and allows to obtain some novel results as well.

\subsection{The Dzyaloshinskii constant with double vector product}

\begin{figure}\includegraphics[width=8cm,angle=0]{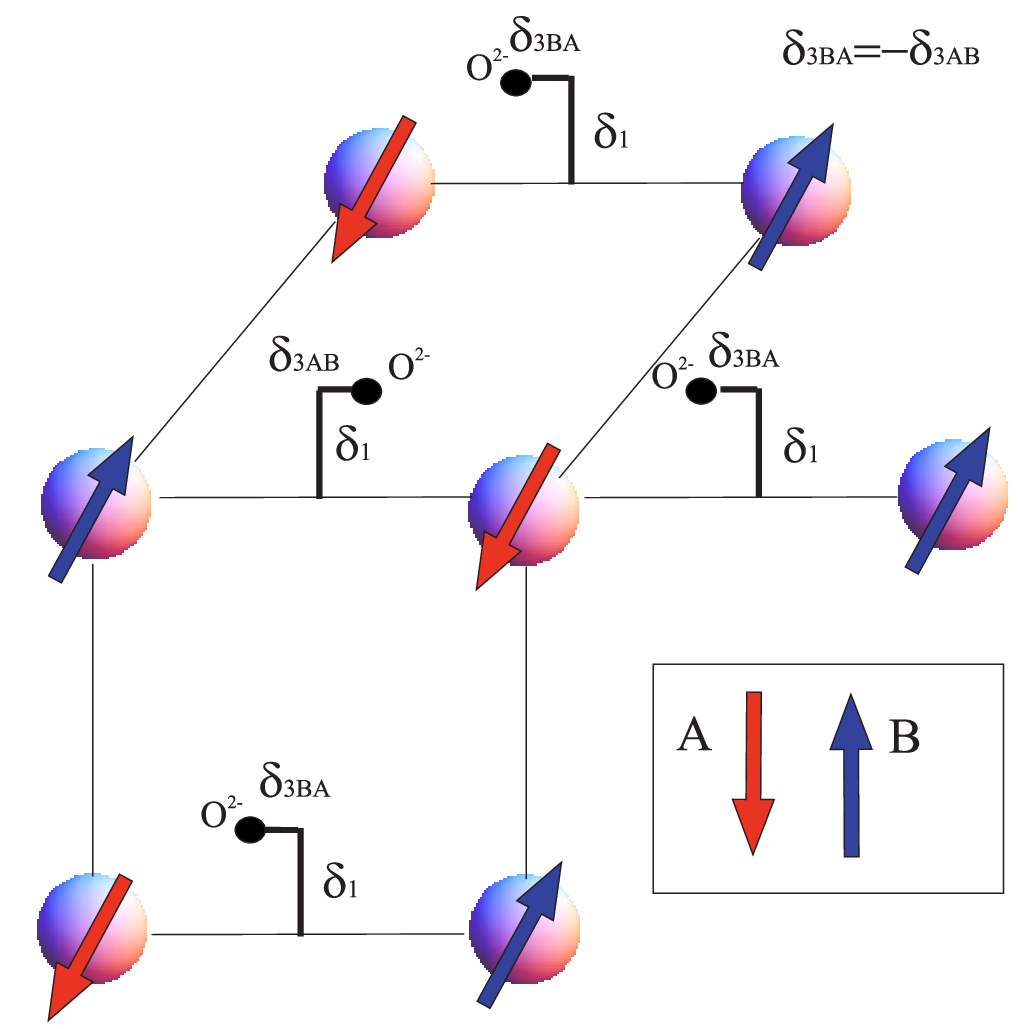}
\caption{\label{MFMextremumP Fig 03} The figure shows
a possibility for the location of the ligand ions relatively magnetic ions
in the antiferromagnetic materials.
This is a modification of the configuration shown in Fig. (\ref{MFMextremumP Fig 02}).
}
\end{figure}

\begin{figure}\includegraphics[width=8cm,angle=0]{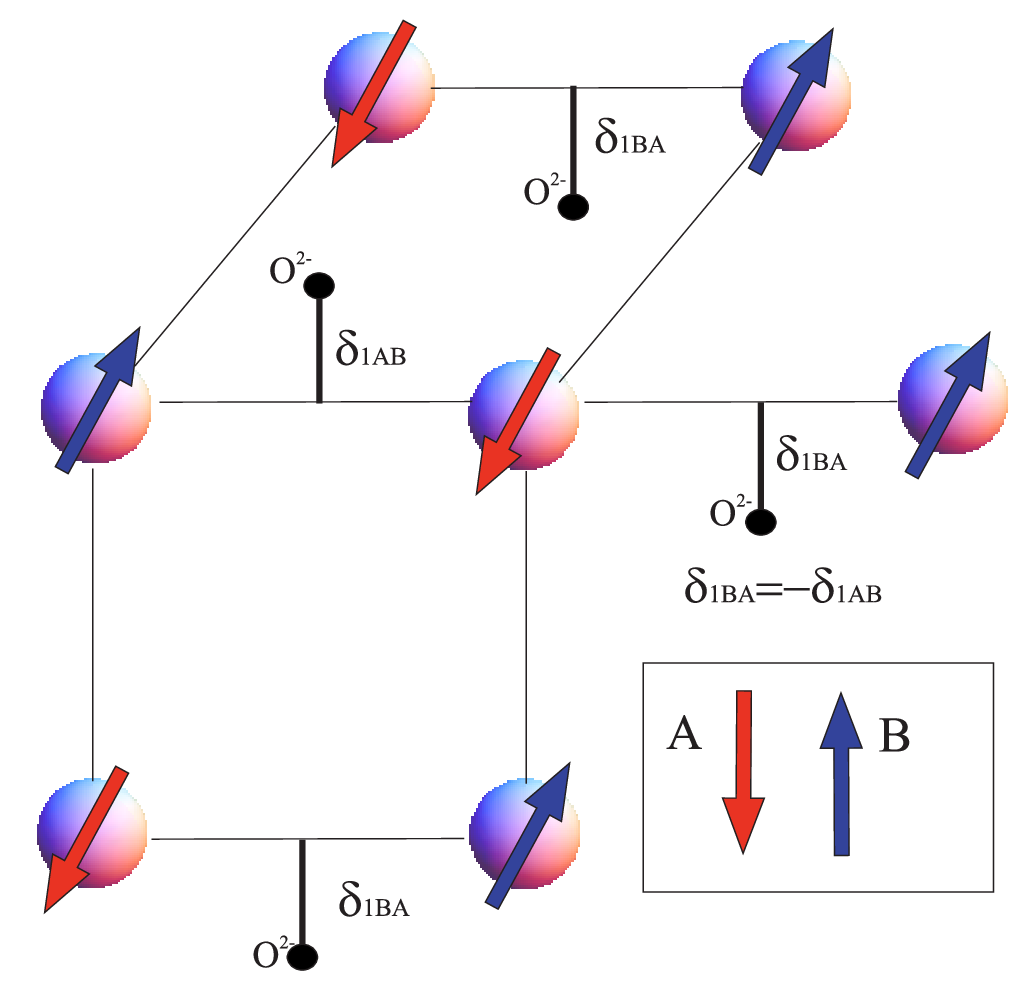}
\caption{\label{MFMextremumP Fig 04} The figure shows
a possibility for the location of the ligand ions relatively magnetic ions
in the antiferromagnetic materials, more complex in comparison with Fig. (\ref{MFMextremumP Fig 01}).
It seems possible to consider the regime given in this figure with the additional shifts of the ligands
$\mbox{\boldmath $\delta$}_{2,ij-AB}$ and $\mbox{\boldmath $\delta$}_{3,ij-AB}$
demonstrated in Figs. (\ref{MFMextremumP Fig 02}) and (\ref{MFMextremumP Fig 03}).
}
\end{figure}

In this section we consider the Dzyaloshinskii constant proportional to the double vector product in more details.
It is presented in equation (\ref{MFMemf D Dz const str G}) as the general anzatz
\begin{equation}\label{MFMemf D Dz const str G double vp 1}
\textbf{D}_{ij,(dvp)}=
\tilde{\zeta}_{2}(r_{ij})[[\textbf{r}_{ij}\times\mbox{\boldmath $\delta$}_{n,ij-AB}]\times\mbox{\boldmath $\delta$}_{n',ij-AB}], \end{equation}
The ligand shift $\mbox{\boldmath $\delta$}$
can be constructed of the partial shifts illustrated in Figs. \ref{MFMextremumP Fig 02}, \ref{MFMextremumP Fig 03}, and \ref{MFMextremumP Fig 04}.
We need to consider possible combinations of these elements for the construction of DMC (\ref{MFMemf D Dz const str G double vp 1}).
Possible combinations are given by four terms
$$\textbf{D}_{ij,(dvp)}=\zeta_{2}(r_{ij})[[\textbf{r}_{ij}\times\mbox{\boldmath $\delta$}_{1-AB}]\times\mbox{\boldmath $\delta$}_{3-AB}]$$
$$+\zeta_{2}'(r_{ij})[[\textbf{r}_{ij}\times\mbox{\boldmath $\delta$}_{2-AB}]\times\mbox{\boldmath $\delta$}_{3-AB}]$$
$$+\zeta_{2}''(r_{ij})[[\textbf{r}_{ij}\times\mbox{\boldmath $\delta$}_{1}]\times\mbox{\boldmath $\delta$}_{1}]$$
\begin{equation}\label{MFMemf D Dz const str G double vp}
+\zeta_{2}'''(r_{ij})[[\textbf{r}_{ij}\times\mbox{\boldmath $\delta$}_{1-AB}]\times\mbox{\boldmath $\delta$}_{1-AB}], \end{equation}
where the subindex $dvp$ stands for "double vector product" to distinguish this part of the Dzyaloshinskii constant.

Main interest is focused on the first term.
This regime can be considered as combination of Figs. \ref{MFMextremumP Fig 03} and \ref{MFMextremumP Fig 04}.
So, there is "simultaneous" "oscillations" of $\mbox{\boldmath $\delta$}_{1-AB}$ presented in Fig. \ref{MFMextremumP Fig 04}
and "oscillations" of $\mbox{\boldmath $\delta$}_{3-AB}$ presented in Fig. \ref{MFMextremumP Fig 03}
(no fixed $\mbox{\boldmath $\delta$}_{1}$ like in Figs. \ref{MFMextremumP Fig 01}, and \ref{MFMextremumP Fig 03} is assumed here).
It can be represented as
$\textbf{D}_{ij,(dvp)}=\zeta_{2}(r_{ij})\mbox{\boldmath $\delta$}_{1-AB}(\textbf{r}_{ij}\cdot\mbox{\boldmath $\delta$}_{3-AB})$
since $(\textbf{r}_{ij}\cdot\mbox{\boldmath $\delta$}_{1-AB})=0$.
Here the Dzyaloshinskii constant is parallel to $\mbox{\boldmath $\delta$}_{1-AB}$ as it is considered in the previous subsection.
But it also depends on the relative distance of the magnetic ions $\textbf{r}_{ij}$.
Hence, it leads to different macroscopic behavior.

The second term requires more complex configuration for its realization.
It requires "simultaneous" existence of $\mbox{\boldmath $\delta$}_{2-AB}$ and $\mbox{\boldmath $\delta$}_{3-AB}$
presented in Figs. \ref{MFMextremumP Fig 02}, and \ref{MFMextremumP Fig 03}
together with fixed $\mbox{\boldmath $\delta$}_{1}$
which is also presented in these figures.
It can be represented as
$\textbf{D}_{ij,(dvp)}=
\zeta_{2}'(r_{ij})(\mbox{\boldmath $\delta$}_{2-AB}(\textbf{r}_{ij}\cdot\mbox{\boldmath $\delta$}_{3-AB})$
since $(\textbf{r}_{ij}\cdot\mbox{\boldmath $\delta$}_{2-AB})=0$.
So, it gives similar to the first term macroscopic terms with corresponding representation of the ligands shifts
$\mbox{\boldmath $\delta$}_{2-AB}$ and $\mbox{\boldmath $\delta$}_{3-AB}$.

Two last terms in equation (\ref{MFMemf D Dz const str G double vp}) are parallel to $\textbf{r}_{ij}$.
They give the renormalization of constant $\gamma_{ij}$ in the Dzyaloshinskii constant (\ref{MFMemf D Dz const str 0}).

\section{Spin density evolution equations}

In this section we consider the contributions of the Dzyaloshinskii-Moriya interaction
in the macroscopic Landau--Lifshitz--Gilbert equation,
corresponding to the different parts of the Dzyaloshinskii constant.

The spin density for the subsystem can be considered in following form
\cite{AndreevTrukh PS 24},
\cite{MaksimovTMP 2001},
\cite{Andreev PoF 21}
\begin{equation}\label{MFMemf spin A def}
\textbf{S}_{A}=\int dR\sum_{i\in A}
\delta(\textbf{r}-\textbf{r}_{i})
\Psi^{\dag}(R,t)\hat{\textbf{S}}_{i}\Psi(R,t).
\end{equation}
This definition allows to derive the spin evolution equation
(or, other terms, the Landau--Lifshitz--Gilbert equation).
Some methodological elements of derivation can be found in Ref. \cite{Andreev LP 21 fermions}.

Equation (\ref{MFMemf spin A def}) contains the following notations:
the LLG equation is formulated for the spin density vector field $\textbf{S}(\textbf{r},t)$,
which is the vector function of space coordinates and time,
it is defined via the many-particle wave function (wave spinor) of all magnetic ions
$\Psi_{S}(R,t)=\Psi_{S}(\textbf{r}_{1}, ..., \textbf{r}_{i}, ..., \textbf{r}_{N},t)$
so $R=\{\textbf{r}_{1}, ..., \textbf{r}_{i}, ..., \textbf{r}_{N}\}$ is the vector in $3N$ dimensional space
showing the aggregate of coordinates of all magnetic ions,
$N$ is the number of the magnetic ions in the system,
$S=\{s_{1}, ..., s_{i}, ..., s_{N}\}$ is the aggregate of spin indexes of all magnetic ions,
$\Psi_{S}^{\dagger}$ is the Hermitian conjugated wave spinor,
$\int ... dR$ is the integral in $3N$ dimensional space,
with $dR=d^{3}r_{1}...d^{3}r_{N}$,
$\hat{\textbf{s}}_{i}$ is the spin operator of $i$-th magnetic ion,
and
$\delta(\textbf{r}-\textbf{r}_{i})$ is the delta-function.

\subsection{Keffer form of the Dzyaloshinskii constant}

Let us present the spin balance equation for the ferromagnetic materials,
where the spin torque is created by the Dzyaloshinskii-Moriya interaction
with the Dzyaloshinskii constant composed of two terms (\ref{MFMemf D Dz const str 0}).
Hence we have two group of terms distinguishable by coefficients
(the interaction constants)
$$\partial_{t}\textbf{S}_{DMI}=
\frac{1}{3}g_{(\gamma)}[\textbf{S}\times curl\textbf{S}]$$
\begin{equation}\label{MFMemf s evolution MAIN TEXT}
+\frac{1}{3}g_{(\beta)}\biggl((\textbf{S}\cdot[\mbox{\boldmath $\delta$}\times\nabla])\textbf{S}
-\frac{1}{2}[\mbox{\boldmath $\delta$}\times\nabla]S^{2}\biggr),
\end{equation}
where we have the following interaction constants related to functions
$\gamma(r_{ij})$ and $\beta(r_{ij})$ in the Dzyaloshinskii constant (\ref{MFMemf D Dz const str 0}):
$g_{(\gamma)}=\int r^2 \gamma(r)d^{3}r$,
and
$g_{(\beta)}=\int r^2 \beta(r)d^{3}r$.

Let us present the similar result for the two component antiferromagnetic materials
with partial spin densities $\textbf{S}_{A}$ and $\textbf{S}_{B}$
leading to collective functions
$\textbf{L}=\textbf{S}_{A}-\textbf{S}_{B}$
and
$\textbf{M}=\textbf{S}_{A}+\textbf{S}_{B}$.
These equations can be written in the following form
$$\partial_{t}\textbf{L}=\frac{1}{3}g_{(\gamma)}[\textbf{M}\times curl\textbf{L}]$$
\begin{equation}\label{MFMemf L evolution MAIN TEXT}
+\frac{1}{3}g_{(\beta)}\biggl((\textbf{M}\cdot[\mbox{\boldmath $\delta$}\times\nabla])\textbf{L}
-M^{\beta}[\mbox{\boldmath $\delta$}\times\nabla]L^{\beta}\biggr),\end{equation}
and
$$\partial_{t}\textbf{M}=\frac{1}{3}g_{(\gamma)}[\textbf{L}\times curl\textbf{L}]$$
\begin{equation}\label{MFMemf M evolution MAIN TEXT}
+\frac{1}{3}g_{(\beta)}\biggl((\textbf{L}\cdot[\mbox{\boldmath $\delta$}\times\nabla])\textbf{L}
-\frac{1}{2}[\mbox{\boldmath $\delta$}\times\nabla]L^{2}\biggr).\end{equation}
Presented torque can be considered as parts of the Lifshitz invariants.

\subsection{Spin density for the Dzyaloshinskii constant parallel to the ligand shift for AFM}

Here we consider the Dzyaloshinskii constant parallel to the ligand shifts.
Fig. \ref{MFMextremumP Fig 02} illustrates the regime considered by
Dzyaloshinskii
in order to describe the reverse magneto-electric effect in the antiferromagnetic material.
Hence, we need to dial with $\mbox{\boldmath $\delta$}_{2,AB}$.
However, the derivation can be done in the same way for any of three types of "oscillating" shifts
$\mbox{\boldmath $\delta$}_{n,AB}$ with $n=1,2,3$.
Other forms of $\mbox{\boldmath $\delta$}_{n,AB}$ are presented in Figs. \ref{MFMextremumP Fig 03} and \ref{MFMextremumP Fig 04}.
Hence, we consider the Dzyaloshinskii constant in the following form
\begin{equation}\label{MFMemf D Dz const str G for SEE}
\textbf{D}_{ij}=\zeta_{1}(r_{ij})\mbox{\boldmath $\delta$}_{n,ij-AB}.
\end{equation}

Here, we present corresponding spin torques calculated up to the second order expansion on the relative distance
$$\partial_{t}\textbf{S}_{A,DMI,2}$$
\begin{equation}\label{MFMemf s evolution MAIN TEXT 2 A}
=
-\biggl[\textbf{S}_{A}\times \biggl[\mbox{\boldmath $\delta$}_{2,AB} \times \biggl(g_{(0\zeta_{1})}\textbf{S}_{B}+\frac{1}{6}g_{(2\zeta_{1})}\triangle\textbf{S}_{B}\biggr)\biggr]\biggr],
\end{equation}
and
$$\partial_{t}\textbf{S}_{B,DMI,2}$$
\begin{equation}\label{MFMemf s evolution MAIN TEXT 2 B}
=
\biggl[\textbf{S}_{B}\times \biggl[\mbox{\boldmath $\delta$}_{2,AB} \times \biggl(g_{(0\zeta_{1})}\textbf{S}_{A}+\frac{1}{6}g_{(2\zeta_{1})}\triangle\textbf{S}_{A}\biggr)\biggr]\biggr],
\end{equation}
while usually the first term in each equation is used.

Next, we also present corresponding torques in equations for the "antiferromagnetic vectors" $\textbf{L}$ and $\textbf{M}$.
First we demonstrate equation for $\textbf{L}=\textbf{S}_{A}-\textbf{S}_{B}$:
$$\partial_{t}\textbf{L}_{2}=\frac{1}{2}g_{(0\zeta_{1})}
\biggl([\textbf{L}\times [\mbox{\boldmath $\delta$}_{n,AB}\times\textbf{L}]]
-[\textbf{M}\times [\mbox{\boldmath $\delta$}_{n,AB}\times\textbf{M}]]\biggr)$$
$$+\frac{1}{12}g_{(2\zeta_{1})}\biggl(\mbox{\boldmath $\delta$}_{n,AB} (\textbf{L}\cdot\triangle\textbf{L})
-\mbox{\boldmath $\delta$}_{n,AB}(\textbf{M}\cdot\triangle\textbf{M})$$
\begin{equation}\label{MFMemf L evolution MAIN TEXT 2}
+(\textbf{M}\cdot\mbox{\boldmath $\delta$}_{n,AB})\triangle\textbf{M}
-(\textbf{L}\cdot\mbox{\boldmath $\delta$}_{n,AB})\triangle\textbf{L}\biggr).\end{equation}
Here, the main contribution is given by the first two terms proportional to the interaction constant $g_{(0\zeta)}$ appearing in the zeroth order on the interaction radius.
We find here the term quadratic on vector $\textbf{M}$
in contrast to other forms of the Dzyaloshinskii-Moriya interaction
(or other interactions in
Landau--Lifshitz--Gilbert equation),
like equation (\ref{MFMemf L evolution MAIN TEXT}),
where $\partial_{t}\textbf{L}$ is proportional to some form of product of $\textbf{L}$ and $\textbf{M}$.
Next, we give equation for $\textbf{M}=\textbf{S}_{A}+\textbf{S}_{B}$:
$$\partial_{t}M^{\alpha}_{2}=\frac{1}{2}g_{(0\zeta_{1})}[\mbox{\boldmath $\delta$}_{n,AB}\times[\textbf{M}\times\textbf{L}]]$$
$$+\frac{1}{12}g_{(2\zeta_{1})}\biggl(\mbox{\boldmath $\delta$}_{n,AB} (\textbf{M}\cdot\triangle\textbf{L})
-\mbox{\boldmath $\delta$}_{n,AB}(\textbf{L}\cdot\triangle\textbf{M})$$
\begin{equation}\label{MFMemf M evolution MAIN TEXT 2}
+(\textbf{L}\cdot\mbox{\boldmath $\delta$}_{n,AB})\triangle\textbf{M}
-(\textbf{M}\cdot\mbox{\boldmath $\delta$}_{n,AB})\triangle\textbf{L}
\biggr)
.\end{equation}
Here all terms are proportional to the product of $\textbf{L}$ and $\textbf{M}$.
It has more similarity to $\partial_{t}\textbf{L}$ in equation (\ref{MFMemf L evolution MAIN TEXT}),
than other form of equation for $\partial_{t}\textbf{M}$ (see equation (\ref{MFMemf M evolution MAIN TEXT})),
where $\partial_{t}\textbf{M}$ is quadratic relatively vector $\textbf{L}$.

\subsection{Spin density for the Dzyaloshinskii constant with double vector product}

The fourth form of the Dzyaloshinskii constant appears for the two-component magnetic structures,
such as the antiferromagnetic materials, similarly to the DMC form considered in previous subsection.
It gives a set of torques which differs from the torques presented above.
Here we repeat the structure of the DMC
\begin{equation}\label{MFMemf D Dz const str G double vp LLG}
\textbf{D}_{ij,(dvp)}=\zeta_{2}(r_{ij})\mbox{\boldmath $\delta$}_{1-AB}(\textbf{r}_{ij}\cdot\mbox{\boldmath $\delta$}_{3-AB}),\end{equation}
which is considered in this subsection.

Next, we use Hamiltonian (\ref{MFMemf Ham DMI}) with the DMC (\ref{MFMemf D Dz const str G double vp LLG})
in order to consider the evolution of the partial spin density (\ref{MFMemf spin A def}).
We obtain the following spin torque in tensor form
$\partial_{t}S_{A}^{\mu}=(1/3)\varepsilon^{\mu\nu\sigma}\varepsilon^{\sigma\alpha\beta}
\delta_{1,AB}^{\alpha}\delta_{3,AB}^{\lambda}g_{(2\zeta_{2})}
S_{A}^{\nu}\nabla^{\lambda}S_{B}^{\beta}$.
For the second subsystem
we find
$\partial_{t}S_{B}^{\mu}=\partial_{t}S_{A}^{\mu}(A\leftrightarrow B)$
$=(1/3)\varepsilon^{\mu\nu\sigma}\varepsilon^{\sigma\alpha\beta}
\delta_{1,AB}^{\alpha}\delta_{3,AB}^{\lambda}g_{(2\zeta_{2})}
S_{B}^{\nu}\nabla^{\lambda}S_{A}^{\beta}$.
Next, we present the spin torque for one of subsystems in the vector notations:
$$\partial_{t}\textbf{S}_{A,DMI,3}=
\frac{1}{3}g_{(2\zeta_{2})}
\biggl[\mbox{\boldmath $\delta$}_{1,AB} S_{A}^{\nu}(\mbox{\boldmath $\delta$}_{3,AB} \cdot\nabla)S_{B}^{\nu}$$
\begin{equation}\label{MFMemf s evolution MAIN TEXT 3}
- (\mbox{\boldmath $\delta$}_{1,AB}\cdot\textbf{S}_{A})(\mbox{\boldmath $\delta$}_{3,AB}\cdot\nabla)\textbf{S}_{B}\biggr].
\end{equation}

We also represent derived torques via the collective antiferromagnetic vectors:
$$\partial_{t}\textbf{L}_{DMI,3}=
\frac{1}{6}g_{(2\zeta_{2})}
\biggl[\mbox{\boldmath $\delta$}_{1,AB} L^{\nu}(\mbox{\boldmath $\delta$}_{3,AB} \cdot\nabla)M^{\nu}$$
$$- (\mbox{\boldmath $\delta$}_{1,AB}\cdot\textbf{L})(\mbox{\boldmath $\delta$}_{3,AB}\cdot\nabla)\textbf{M}
-\mbox{\boldmath $\delta$}_{1,AB} M^{\nu}(\mbox{\boldmath $\delta$}_{3,AB} \cdot\nabla)L^{\nu}$$
\begin{equation}\label{MFMemf L evolution MAIN TEXT 3}
+(\mbox{\boldmath $\delta$}_{1,AB}\cdot\textbf{M})(\mbox{\boldmath $\delta$}_{3,AB}\cdot\nabla)\textbf{L}\biggr],\end{equation}
and
$$\partial_{t}\textbf{M}_{DMI,3}=
\frac{1}{6}g_{(2\zeta_{2})}
\biggl[-\frac{1}{2}\mbox{\boldmath $\delta$}_{1,AB} (\mbox{\boldmath $\delta$}_{3,AB} \cdot\nabla)L^{2}$$
$$+(\mbox{\boldmath $\delta$}_{1,AB}\cdot\textbf{L})(\mbox{\boldmath $\delta$}_{3,AB}\cdot\nabla)\textbf{L}
+\frac{1}{2}\mbox{\boldmath $\delta$}_{1,AB} (\mbox{\boldmath $\delta$}_{3,AB} \cdot\nabla)M^{2}$$
\begin{equation}\label{MFMemf M evolution MAIN TEXT 3}
-(\mbox{\boldmath $\delta$}_{1,AB}\cdot\textbf{M})(\mbox{\boldmath $\delta$}_{3,AB}\cdot\nabla)\textbf{M}\biggr].\end{equation}

Each of equations (\ref{MFMemf s evolution MAIN TEXT 3}), (\ref{MFMemf L evolution MAIN TEXT 3}), and (\ref{MFMemf M evolution MAIN TEXT 3}) has two group of terms
(in relation to the direction):
one proportional to the partial ligand shift
$\mbox{\boldmath $\delta$}_{1,AB}$,
and the second group
is directed along the spin density
being under derivatives
($\textbf{S}_{B}$ in equation for $\textbf{S}_{A}$,
both $\textbf{L}$ and $\textbf{M}$ in equations for the antiferromagnetic vectors).

\begin{figure}\includegraphics[width=8cm,angle=0]{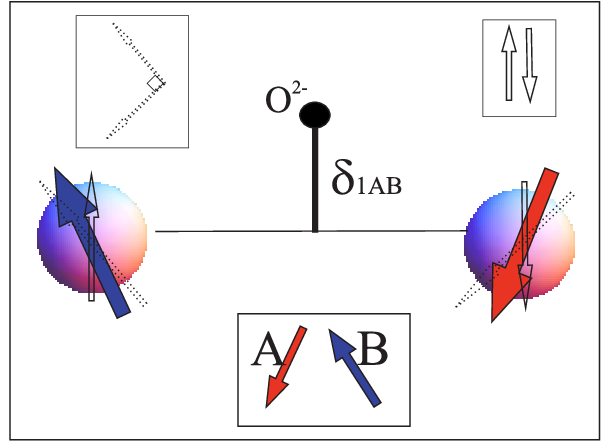}
\caption{\label{MFMextremumP Fig 05}
The figure illustrates configuration of spins with an nonzero angle between approximately antiparallel spins.
Hence, both mechanisms of the polarization formation described in Sec. V can be realized. }
\end{figure}

\section{Energy density}

Construction and application of the free energy density is the common method of analysis of the magnetic phenomenon.
However, we derive the spin evolution equations presented above directly from quantum mechanics.
Nevertheless, it can be useful to present the energy density for the
Dzyaloshinskii-Moriya interaction
with chosen forms of the Dzyaloshinskii constants.

To this end, we need to present the definition of energy density in terms of wave function.
It is useful to introduce the partial energy densities for each subsystem:
$$\mathcal{E}_{A}=-\frac{1}{2}
\int dR\sum_{i\in A, j\in B}
\delta(\textbf{r}-\textbf{r}_{i})\times$$
\begin{equation}\label{MFMemf enrgy definition}
\times\Psi^{\dag}(R,t)(\textbf{D}_{ij}\cdot[\hat{\textbf{S}}_{i}\times \hat{\textbf{S}}_{j}])\Psi(R,t).
\end{equation}
and
$$\mathcal{E}_{B}=-\frac{1}{2}
\int dR\sum_{i\in A, j\in B}
\delta(\textbf{r}-\textbf{r}_{j})\times$$
\begin{equation}\label{MFMemf enrgy definition B}
\times\Psi^{\dag}(R,t)(\textbf{D}_{ij}\cdot[\hat{\textbf{S}}_{i}\times \hat{\textbf{S}}_{j}])\Psi(R,t).
\end{equation}
where the full energy is the sum of partial energies
$\mathcal{E}=\mathcal{E}_{A}+\mathcal{E}_{B}$.
Expressions (\ref{MFMemf enrgy definition}) and (\ref{MFMemf enrgy definition B}) depend on the form of the Dzyaloshinskii constant,
so it gives different expressions.
Notations in equations (\ref{MFMemf enrgy definition}) and (\ref{MFMemf enrgy definition B}) are similar to equation (\ref{MFMemf spin A def}).

\emph{Regimes 1-2.}
First we present the energy density corresponding to the Dzyaloshinskii constant of form of
(\ref{MFMemf D Dz const str 0}):
$\mathcal{E}_{A}=(1/6)\varepsilon^{\alpha\beta\gamma}(g_{(2\gamma)}S_{A}^{\beta}\partial^{\alpha}S_{B}^{\gamma}$
$+g_{(2\beta)}\varepsilon^{\alpha ab}\delta_{1}^{b} S_{A}^{\beta}\partial^{a}S_{B}^{\gamma})$,
and
$\mathcal{E}_{B}=(1/6)\varepsilon^{\alpha\beta\gamma}(g_{(2\gamma)}S_{B}^{\beta}\partial^{\alpha}S_{A}^{\gamma}$
$+g_{(2\beta)}\varepsilon^{\alpha ab}\delta_{1}^{b} S_{B}^{\beta}\partial^{a}S_{A}^{\gamma})$.
We also present one of partial energy densities in vector notations:
$$\mathcal{E}_{A}=-\frac{1}{6}
g_{(2\gamma)}(\textbf{S}_{A}\cdot [\nabla\times\textbf{S}_{B}])$$
\begin{equation}\label{MFMemf energy fin 2 A}
+\frac{1}{6}g_{(2\beta)}(\textbf{S}_{A}\cdot[[\mbox{\boldmath $\delta$}_{1}\times\nabla]\times\textbf{S}_{B}]).
\end{equation}
For the full energy density we use notations of collective function of the antiferromagnetic vectors:
$$\mathcal{E}=\mathcal{E}_{A}+\mathcal{E}_{B}=-\frac{1}{12}
g_{(2\gamma)}\{(\textbf{M}\cdot [\nabla\times\textbf{M}])-(\textbf{L}\cdot [\nabla\times\textbf{L}])\}$$
\begin{equation}\label{MFMemf energy fin 2 LM}
+\frac{1}{12}g_{(2\beta)}\{\textbf{M}\cdot[[\mbox{\boldmath $\delta$}_{1}\times\nabla]\times\textbf{M}]
-\textbf{L}\cdot[[\mbox{\boldmath $\delta$}_{1}\times\nabla]\times\textbf{L}]\}.
\end{equation}

\emph{Regimes 3.}
Next, we find the energy density for the DMI with DMC given by equation (\ref{MFMemf D Dz const str G par delta}),
we also repeat it here:
$\textbf{D}_{ij}=\zeta_{1}(r_{ij})\mbox{\boldmath $\delta$}_{2,ij-AB}$.

Hence,
we obtain the following expression considered up to the second order of the expansion on the relative distance of the magnetic ions
$$\mathcal{E}=-\mbox{\boldmath $\delta$}_{2,AB}\cdot\biggl[g_{(0\zeta_{1})} [\textbf{S}_{A}\times\textbf{S}_{B}] $$
\begin{equation}\label{MFMemf energy fin 3}
+\frac{1}{12}g_{(2\zeta_{1})}\biggl([(\triangle\textbf{S}_{A})\times\textbf{S}_{B}]+[\textbf{S}_{A}\times(\triangle\textbf{S}_{B})]\biggr)\biggr].
\end{equation}
Up to the notations we get the energy density suggested
by Dzyaloshinskii
$\mathcal{E}_{0}=\tilde{\textbf{n}}\cdot [\textbf{S}_{A}\times\textbf{S}_{B}]$
with $\tilde{\textbf{n}}=-g_{(0\zeta_{1})}\mbox{\boldmath $\delta$}_{2,AB}$.
\textit{This conclusion is an essential part of this work
since it shows that chosen form of constant} (\ref{MFMemf D Dz const str G par delta})
\textit{represent required interaction}.

\emph{Regimes 4.}
The fourth form of DMC leads to the following partial energy densities:
$\mathcal{E}_{A}= (1/6)g_{(2\zeta_{2})}\varepsilon^{\alpha\beta\gamma}\delta^{\alpha}_{1,AB}\delta^{\lambda}_{3,AB}S_{A}^{\beta}\partial^{\lambda} S_{B}^{\gamma}$,
and
$\mathcal{E}_{B}= \mathcal{E}_{A}(A\leftrightarrow B)=(1/6)g_{(2\zeta_{2})}\varepsilon^{\alpha\beta\gamma}\delta^{\alpha}_{1,AB}\delta^{\lambda}_{3,AB}S_{B}^{\beta}\partial^{\lambda} S_{A}^{\gamma}$.
We also represent one of them in the vector notations:
\begin{equation}\label{MFMemf energy fin 4 SaSb}
\mathcal{E}_{A}=\frac{1}{6}g_{(2\zeta_{2})}(\mbox{\boldmath $\delta$}_{1,AB}\cdot[\textbf{S}_{A}\times((\mbox{\boldmath $\delta$}_{3,AB}\cdot\nabla)\textbf{S}_{B})]).
\end{equation}
Here, the space derivative acting on the spin density $\textbf{S}_{B}$ is directed parallel to the third partial ligand shift $\mbox{\boldmath $\delta$}_{3,AB}$,
while all combination of spin densities is projected on the direction of "oscillating" first partial ligand shift $\mbox{\boldmath $\delta$}_{1,AB}$.

Full energy density is also presented for this regime,
but in terms of the collective antiferromagnetic vectors:
$$\mathcal{E}=\mathcal{E}_{A}+\mathcal{E}_{B}
=\frac{1}{12}g_{(2\zeta_{2})}
\biggl((\mbox{\boldmath $\delta$}_{1,AB}\cdot[\textbf{M}\times((\mbox{\boldmath $\delta$}_{3,AB}\cdot\nabla)\textbf{M})])$$
\begin{equation}\label{MFMemf energy fin 4 LM}
-(\mbox{\boldmath $\delta$}_{1,AB}\cdot[\textbf{L}\times((\mbox{\boldmath $\delta$}_{3,AB}\cdot\nabla)\textbf{L})])\biggr).
\end{equation}

Combination of equations
(\ref{MFMemf energy fin 2 LM}),
(\ref{MFMemf energy fin 3}), and
(\ref{MFMemf energy fin 4 LM})
shows the full energy density for DMI.

\section{Polarization of the multiferroic material}

There are three regimes of the spin-related polarization formation in the multifferroic materials \cite{Tokura RPP 14}.
Two of them can be modeled via the spin-current model,
where different physical mechanisms are considered for the spin-current appearance.
We focus on these regimes of the polarization described by the spin-current model.
They are described in this section
as the physical background for the derivation of the polarization evolution equation below in this paper.
Corresponding spin orientation is demonstrated in Fig. \ref{MFMextremumP Fig 05},
which allows to manifest both mechanisms described in this section.

The macroscopic form of the spin-current model can be presented within the following equation
\cite{AndreevTrukh JETP 24}, \cite{AndreevTrukh PS 24}, \cite{Tokura RPP 14}, \cite{Mostovoy npj 24} eq. 10:
\begin{equation}\label{MFMemf spin current model} P^{\mu}
=\frac{\gamma}{c}\varepsilon^{\mu\alpha\beta}J^{\alpha\beta},
\end{equation}
where
$J^{\alpha\beta}$ is the spin-current tensor
(the first index is associated with the spin,
and the second index is associated with the momentum),
and $\varepsilon^{\mu\alpha\beta}$ is the Levi-Civita symbol.
Originally, the spin-current model is suggested for the microscopic spin currents forming the electric dipole moment
$d^{\mu}
\sim\varepsilon^{\mu\alpha\beta}J^{\alpha\beta}$
in Ref. \cite{Katsura PRL 05}.

The macroscopic polarization can be defined as
\begin{equation}\label{MFMemf pol A def}
\textbf{P}_{A}=\int dR\sum_{i\in A}
\delta(\textbf{r}-\textbf{r}_{i})
\Psi^{\dag}(R,t)\hat{\textbf{d}}_{i}\Psi(R,t).
\end{equation}
This definition allows to obtain approximate macroscopic polarization.
However, it also allows to derive the polarization evolution equation
(it is also considered
in the next section).
Different forms of the electric dipole moments or macroscopic polarizations are discussed in Refs. \cite{Tokura RPP 14} and \cite{Dong AinP 15}.

\subsection{Regime of "noncollinear" spins}

The electric dipole moment can be formed due to the existence of the noncollinear parts of the neighboring spins.
The electric dipole moment in this regime is proportional to the vector product of two spins.
The vector product change sing at the renumeration of the spins.
However, the electric dipole moment should not depend on the numeration,
hence the electric dipole moment includes the vector of relative position $\textbf{r}_{ij}$ of spins
and it is presented in the following form (see for instance Ref. \cite{Mostovoy npj 24} eq. 10)
\begin{equation}\label{MFMemf edm operator NONsimm}
\hat{\textbf{d}}_{ij}= \alpha_{ij}
[\textbf{r}_{ij}\times[\hat{\textbf{s}}_{i}\times\hat{\textbf{s}}_{j}]],
\end{equation}
where $\alpha_{ij}$ is the coefficient depending on the module of relative position $\mid\textbf{r}_{ij}\mid=r_{ij}$ of spins.

The approximate macroscopic electric polarization for the ferromagnetic multiferroic materials
can be obtained using operator (\ref{MFMemf edm operator NONsimm})
(like in Refs. \cite{AndreevTrukh JETP 24}, \cite{AndreevTrukh PS 24}):
\begin{equation}\label{MFMemf P def expanded noncoll}
\textbf{P}(\textbf{r},t)=
\frac{1}{3}g_{(2\alpha)}
[(\textbf{S}\cdot\nabla)\textbf{S}-\textbf{S}(\nabla\cdot \textbf{S})], \end{equation}
where
$g_{(2\alpha)}=\int \alpha(r)r^{2}d^{3}r$.
Expression (\ref{MFMemf P def expanded noncoll}) can be found in earlier
Refs. \cite{Sparavigna PRB 94}, \cite{Baryakhtar JETP 83}, \cite{Mostovoy PRL 06}.


Similar expressions can be found for the partial polarization in antiferromagnetic materials 
\begin{equation}\label{MFMemf P def expanded noncoll AFM A}
\textbf{P}_{A}(\textbf{r},t)=
\frac{1}{3}g_{(2\alpha,AB)}
[(\textbf{S}_{A}\cdot\nabla)\textbf{S}_{B}-\textbf{S}_{A}(\nabla\cdot \textbf{S}_{B})], \end{equation}
where
$g_{(2\alpha,AB)}=-g_{(2\alpha,AA)}=-g_{(2\alpha,BB)}=-g_{(2\alpha)}$
since $g_{(2\alpha,AB)}$ is proportional to the exchange integral of the symmetric Heisenberg exchange interaction
\cite{AndreevTrukh JETP 24}, \cite{AndreevTrukh PS 24}.
For the second component of polarization we find
$$\textbf{P}_{B}(\textbf{r},t)=\textbf{P}_{A}(A\leftrightarrow B)=$$
\begin{equation}\label{MFMemf P def expanded noncoll AFM B}
=\frac{1}{3}g_{(2\alpha,AB)}
[(\textbf{S}_{B}\cdot\nabla)\textbf{S}_{A}-\textbf{S}_{B}(\nabla\cdot \textbf{S}_{A})]. \end{equation}

Complete polarization can be composed of the partial polarizations and also represented via vectors
$\textbf{L}=\textbf{S}_{A}-\textbf{S}_{B}$
and
$\textbf{M}=\textbf{S}_{A}+\textbf{S}_{B}$:
$$\textbf{P}(\textbf{r},t)=\textbf{P}_{A}+\textbf{P}_{B}=
\frac{1}{6}g_{(2\alpha,AB)}
\biggl[\textbf{L}(\nabla\cdot \textbf{L})$$
\begin{equation}\label{MFMemf P def expanded noncoll AFM AB LM}
-(\textbf{L}\cdot\nabla)\textbf{L}
-\textbf{M}(\nabla\cdot \textbf{M})+(\textbf{M}\cdot\nabla)\textbf{M}\biggr]. \end{equation}
The equations for electric polarization in the antiferromagnetic multiferroic materials are also discussed in Refs.
\cite{Sparavigna PRB 94},
\cite{ZvezdinMukhin JETP L 09},
and
\cite{Podkletnova JETP L 23}.

In order to understand the physical mechanisms behind equations (\ref{MFMemf edm operator NONsimm}) and (\ref{MFMemf P def expanded noncoll})
we need to consider the spin-current model.
Particularly, we follow its form derived from the quantum hydrodynamics \cite{AndreevTrukh JETP 24}, \cite{AndreevTrukh PS 24},
where it is found as result of balance of forces in multiferroic crystal.
The polarization form as the result of balance of the dipole-dipole interaction and the spin-orbit interaction forces.
One of mechanisms of the effective spin current is the magnon spin-current,
which can be found in the spin evolution equation from the torque caused by symmetric Heisenberg exchange interaction.
It shows that constant $\alpha$ in equation (\ref{MFMemf edm operator NONsimm})
is proportional to the exchange integral in symmetric Heisenberg Hamiltonian
\cite{AndreevTrukh JETP 24}, \cite{AndreevTrukh PS 24}.
A similar analysis can be found in review paper \cite{Tokura RPP 14} (see p. 7, Sec. 2.3, between equations 14 and 19),
but the conclusions made there is quite different.

Additional physical mechanism of the polarization formation related to the quantum spin-current associated with the contribution of the quantum Bohm potential is considered in Ref. \cite{AndreevTrukh JETP 24}.
It also work for the "noncollinear" spins and presents the generalization of equation (\ref{MFMemf P def expanded noncoll}).
In the simplest case it gives renormalization of constant $g_{(2\alpha)}$,
while it gives additional dependence on the perturbations of the particle number concentration at the presence of the phonons in the system dynamics.

The contribution of the magnetoelectric effect in
the macroscopic Landau--Lifshitz--Gilbert equation
associated with the polarization (\ref{MFMemf P def expanded noncoll})
can be written via the following spin torque (for the ferromagnetic samples)
\cite{Risinggard SR 16}, \cite{Andreev 2025 05}:
$$\textbf{T}=\frac{1}{3}g_{(2\alpha)}
\biggl[ [\textbf{E}\times \nabla] S^{2}
-2(\textbf{S}\cdot[\textbf{E}\times\nabla]) \textbf{S}$$
\begin{equation}\label{MFMemf Torque magEl nonCol}
-S^{2}(\nabla\times\textbf{E})
+\textbf{S}(\textbf{S}\cdot [\nabla\times\textbf{E}])
\biggr].
\end{equation}

For the antiferromagnetic samples,
we present the spin torque acting on the partial spin density $\textbf{S}_{A}$
$$\textbf{T}_{A}=\frac{1}{3}g_{(2\alpha,AB)}
\biggl[ 2S_{A}^{\nu}[\textbf{E}\times \nabla] S_{B}^{\nu}
-2(\textbf{S}_{A}\cdot[\textbf{E}\times\nabla]) \textbf{S}_{B}$$
\begin{equation}\label{MFMemf Torque magEl nonCol AFM A}
-(\textbf{S}_{A}\cdot\textbf{S}_{B})[\nabla\times\textbf{E}]
+\textbf{S}_{B}(\textbf{S}_{A}
\cdot [\nabla\times\textbf{E}])
\biggr].
\end{equation}
The spin torque acting on the partial spin density $\textbf{S}_{B}$
can be found from equation (\ref{MFMemf Torque magEl nonCol AFM A})
via reorganization of indexes of subspecies $A$ and $B$:
$\textbf{T}_{B}=\textbf{T}_{A}(A\leftrightarrow B)$.

Equation (\ref{MFMemf Torque magEl nonCol AFM A}) can be represented in another equivalent form:
$$\textbf{T}_{A}=\frac{1}{3}g_{(2\alpha,AB)}
\biggl[2E^{\nu}[\textbf{S}_{A}\times \nabla]S_{B}^{\nu}$$
\begin{equation}\label{MFMemf Torque magEl nonCol AFM A2}
-2[\textbf{S}_{A}\times\textbf{E}](\nabla\cdot\textbf{S}_{B})
+S_{B}^{\nu}[\textbf{S}_{A}\times\nabla]E^{\nu} - [\textbf{S}_{A}\times(\textbf{S}_{B}\cdot\nabla)\textbf{E}]
\biggr].
\end{equation}

Equations for the collective antiferromagnetic vectors $\textbf{L}$ and $\textbf{M}$ contain contributions
$\textbf{T}_{A}-\textbf{T}_{B}$ for $\partial_{t}\textbf{L}$
and
$\textbf{T}_{A}+\textbf{T}_{B}$ for $\partial_{t}\textbf{M}$,
which can be rewritten via $\textbf{L}$ and $\textbf{M}$ as well.

\subsection{Regime of "collinear" spins}

One of possibilities of the formation of the electric dipole moment $\textbf{d}$
in the magnetically order material is associated with the collinear parts of the neighboring spins
(or the magnetic moments) \cite{Tokura RPP 14}.
Corresponding relation can be demonstrated for two ions
\begin{equation}\label{MFMemf edm operator simm}
\hat{\textbf{d}}_{ij}= \mbox{\boldmath $\Pi$}_{ij} (\hat{\textbf{s}}_{i}\cdot\hat{\textbf{s}}_{j}).
\end{equation}
The collinear parts of the spins enter this equation via the scalar product of the spin operators
$(\hat{\textbf{s}}_{i}\cdot\hat{\textbf{s}}_{j})$,
while spins generally are noncollinear.
The construction of the electric dipole moment vector $\textbf{d}$ requires introduction of the vector constant
$\mbox{\boldmath $\Pi$}_{ij}$.

Let us discuss the physical mechanisms behind equation (\ref{MFMemf edm operator simm})
following Refs. \cite{AndreevTrukh JETP 24}, \cite{AndreevTrukh PS 24}.
The balance of forces of the electric dipole-dipole interaction and the spin-orbit interaction associated with the motion of the magnetic moments of ions in the electric field of surrounding dipoles leads to the relation between the polarization of medium and the spin-current in the medium
$P^{\mu}=\frac{\gamma}{c}\varepsilon^{\mu\alpha\beta}J^{\alpha\beta}$.
This relation is also called the spin-current model of multiferroics of spin origin \cite{Tokura RPP 14}.
If one considers the effective spin current caused by the
the Dzyaloshinskii-Moriya interaction
with the Dzyaloshinskii constant in the Keffer form
$\textbf{D}_{ij}=\beta(r_{ij})[\textbf{r}_{ij}\times\mbox{\boldmath $\delta$}_{1}]$
\cite{Khomskii JETP 21}, \cite{Fishman PRB 19},
related to the shift of the ligand ion
$\mbox{\boldmath $\delta$}_{1}$,
this spin current can be placed in the spin-current model to find corresponding polarization:
\begin{equation}\label{MFMemf P appr Symm}
\textbf{P}(\textbf{r},t)= \mbox{\boldmath $\delta$}_{1}
[ c_{0}(\textbf{S}\cdot\textbf{S})+c_{2}(\textbf{S}\cdot\triangle\textbf{S})]
, \end{equation}
where coefficients $c_{0}$ and $c_{2}$ are related to the Dzyaloshinskii constant
(it is demonstrated below).
Let us point out that the
effective spin current caused by the
the Dzyaloshinskii-Moriya interaction
is not related to the motion of particles in space,
but it can be considered as a part of the spin-current associated with magnons
and the spin transfer at the spin wave propagation.

Moreover, we can consider
the effective spin current caused by the
the Dzyaloshinskii-Moriya interaction
with the Dzyaloshinskii constant in the form of
$\textbf{D}_{ij}=\gamma(r_{ij})\textbf{r}_{ij}$
(see \cite{Fishman PRB 19}),
but it leads to the symmetric spin-current tensor $J^{\alpha\beta}_{(\gamma)}=J^{\beta\alpha}_{(\gamma)}$
and the zero polarization in accordance with the spin-current model
$P^{\mu}=\frac{\gamma}{c}\varepsilon^{\mu\alpha\beta}J^{\alpha\beta}$.
The application of the quantum hydrodynamic method to the electric dipole moment (\ref{MFMemf edm operator simm})
allows to obtain same expression for the macroscopic polarization as derived from the spin-current model
(\ref{MFMemf P appr Symm}).
Hence, it gives relation of vector coefficient $\mbox{\boldmath $\Pi$}_{ij}$
in the electric dipole moment (\ref{MFMemf edm operator simm}) with the elements of
the Dzyaloshinskii constant in the Keffer form
$\textbf{D}_{ij}=\beta(r_{ij})[\textbf{r}_{ij}\times\mbox{\boldmath $\delta$}_{1}]$.
These relations have the following form \cite{AndreevTrukh PS 24}:
$c_{0}\mbox{\boldmath $\delta$}_{1}=\int \mbox{\boldmath $\Pi$}_{ij}(r_{ij})d^{3}r_{ij}$,
$c_{2}\mbox{\boldmath $\delta$}_{1}=(1/6)\int r_{ij}^{2}\mbox{\boldmath $\Pi$}_{ij}(r_{ij})d^{3}r_{ij}$,
and
$\frac{\partial \mbox{\boldmath $\Pi$}}{\partial r}
=\frac{\gamma}{c}r\beta(r)\mbox{\boldmath $\delta$}_{1}$.
The last equation can be considered in different form:
$\mbox{\boldmath $\Pi$}
=-\frac{\gamma}{3c}r^2\beta(r)\mbox{\boldmath $\delta$}_{1}$,
but this representation is possible for the zero order of the expansion proportional to $c_{0}$ only.


Following Ref. \cite{AndreevTrukh PS 24}
we can present the generalization of polarization (\ref{MFMemf P appr Symm})
on the antiferromagnetic multiferroic materials:
\begin{equation}\label{MFMafmUEI P def expanded Symm AFR}
\textbf{P}=\mbox{\boldmath $\delta$}_{1}[2 c_{0,AB}
(\textbf{S}_{A}\cdot\textbf{S}_{B})
+c_{2,AB}
(S^{\nu}_{A} \triangle S^{\nu}_{B} +S^{\nu}_{B} \triangle S^{\nu}_{A})].
\end{equation}
The polarization is related to the shift of ion located between magnetic ions with opposite spins.
Formally, we can consider partial polarizations $\textbf{P}_{A}$ and $\textbf{P}_{B}$ in accordance with equation (\ref{MFMemf pol A def}),
so $\textbf{P}=\textbf{P}_{A}+\textbf{P}_{B}$.
Hence, one of partial polarizations is
$\textbf{P}_{A}=\mbox{\boldmath $\delta$}_{1}
[ c_{0,AB}
(\textbf{S}_{A}\cdot\textbf{S}_{B})
+c_{2,AB}
(\textbf{S}_{A}\cdot \triangle \textbf{S}_{B})] $.

Let us also present the spin-torque related to the magneto-electric effect described in this subsection
\cite{Andreev 25 10}:
$$\textbf{T}=\partial_{t}\textbf{S}=
2 c_{2}\biggl((\mbox{\boldmath $\delta$}\cdot \textbf{E})
\cdot[\textbf{S}\times \triangle \textbf{S}]$$
\begin{equation}\label{MFMemf Torque magEl Col FM}
+(\mbox{\boldmath $\delta$}\cdot(\partial^{\nu} \textbf{E}))\cdot[\textbf{S}\times\partial^{\nu} \textbf{S}]\biggr).
\end{equation}
This expression is given for the ferromagnetic materials.

In Ref.
\cite{Andreev 25 10}
the spin torque related to the magnetoelectric effect in the antiferromagnetic multiferroics with contribution of "collinear" parts of spins:
$$\textbf{T}_{L}=\partial_{t}\textbf{L}=2 c_{0}
(\mbox{\boldmath $\delta$}\cdot \textbf{E}) \cdot[\textbf{M}\times\textbf{L}]
$$
\begin{equation}\label{MFMemf Torque magEl Col AFM}
+2 c_{2}
[(\mbox{\boldmath $\delta$}\cdot \textbf{E})\cdot[\textbf{M}\times\triangle\textbf{L}]
+(\mbox{\boldmath $\delta$}\cdot(\partial^{\nu} \textbf{E}))\cdot[\textbf{M}\times\partial^{\nu}\textbf{L}] ],
\end{equation}
and
$$\textbf{T}_{M}=\partial_{t}\textbf{M}=2 c_{2}
[(\mbox{\boldmath $\delta$}\cdot \textbf{E})\cdot[\textbf{L}\times\triangle \textbf{L}]
$$
\begin{equation}\label{MFMemf Torque magEl Col AFM}
+(\mbox{\boldmath $\delta$}\cdot(\partial^{\nu} \textbf{E}))\cdot[\textbf{L}\times\partial^{\nu} \textbf{L}]].
\end{equation}
Let us to point out that equation (\ref{MFMemf Torque magEl Col AFM})
contains a term (the first term, proportional to $c_{0}$)
which has no analogue in the ferromagnetic limit.

\subsection{Discussion for the polarization mechanisms}

Interesting behavior demonstrates the polarization of the multiferroic materials in the considered regimes.
It is demonstrated
that
the electric dipole moment associated with parallel parts of spins (\ref{MFMemf edm operator simm})
appears due to the Dzyaloshinskii-Moriya interaction,
while the Dzyaloshinskii-Moriya interaction exists if spins are noncollinear.
In reverse,
the electric dipole moment associated with the perpendicular parts of spins (\ref{MFMemf edm operator NONsimm})
is caused by the symmetric Heisenberg exchange interaction,
while this interaction exists for the collinear spins and tends to make spins collinear.

Physical picture can be described in the following way.
No matter which interactions
(it can be the Dzyaloshinskii-Moriya interaction)
creates the non-collinear order of spins in the system
(usually some small deviation from the collinear order for the nearest spins).
These interactions compete with the symmetric Heisenberg exchange interaction,
which tends to create the collinear order.
As the result, some partially collinear order is formed.
This can be considered as the first level of the hierarchy of the formation of the equilibrium state in the medium,
which is related to the form of the equilibrium spin density
(it can be found from the minimization of the energy density or the balance of the torques in the spin evolution equations).
As the second level of the hierarchy we can consider formation of the polarization on the found spin structure.
The polarization is related to the force acting on the charges created by the particular spin configuration.
The effective spin-current induces the force field acting on the charges related to the spin-orbit interaction.
On one hand, the spin-orbit interaction shows the action of the electric field (or the charge) on the moving magnetic moment.
In reverse, the spin-orbit interaction shows the action of the moving magnetic moment or the effective spin-currents on the charges.
In reality all described effects act simultaneously and the polarization affects the equilibrium spin structure,
but this schematic picture allows to understand that there is no similarity between the form of the Hamiltonian of an interaction and the form of the electric dipole moment which is formed due to this interaction.


\section{Polarization evolution equation}

The polarization evolution equation in the multiferroic materials is considered in literature in few cases:
the symmetric Heisenberg Hamiltonian for the "collinear" spins \cite{Andreev 23 12},
the symmetric Heisenberg Hamiltonian for the "non-collinear" spins \cite{AndreevTrukh JETP 24},
and
the Zeeman energy for the "non-collinear" spins \cite{Andreev 23 11}.
Definately, the symmetric Heisenberg Hamiltonian and the Zeeman energy are terms
(together with the anisotropy energy)
giving the main contribution in the evolution of magnetic moments and the polarization in magnetically ordered multiferroics.
However, the Dzyaloshinskii-Moriya interaction is an important part of the model of the multiferroics.
Therefore, we present the contribution of novel parts of the Keffer form of Dzyaloshinskii constant in the polarization evolution equation.
Here, we consider both regimes of polarization:
the "collinear" spins and the "non-collinear" spins.

\subsection{Polarization evolution for the Dzyaloshinskii constant parallel to the ligand shift}

Here we consider single regime of interaction, which exists for the AFM,
where
the Dzyaloshinskii constant has the following form
\begin{equation}\label{MFMemf D Dz const str G for PEE}
\textbf{D}_{ij}=\zeta_{1}(r_{ij})\mbox{\boldmath $\delta$}_{2,ij-AB}.
\end{equation}
We consider the polarization evolution equations under influence of the chosen interaction for two types of the electric dipole moments
(\ref{MFMemf edm operator NONsimm}) and (\ref{MFMemf edm operator simm}).

\subsubsection{Regime of "collinear" spins}

In this subsection we take polarization
(\ref{MFMemf pol A def})
with the electric dipole moment (\ref{MFMemf edm operator simm}).
We consider the time derivative of this function and use the non-stationary Schrodinger equation with the Dzyaloshinskii-Moriya Hamiltonian with the Dzyaloshinskii constant (\ref{MFMemf D Dz const str G for PEE}).
First, we derive the polarization evolution equations for the partial polarization $\partial_{t}\textbf{P}_{A}$.
However, further calculations show that
$\partial_{t}\textbf{P}_{B}=\partial_{t}\textbf{P}_{A}=-\partial_{t}\textbf{P}_{A}(A\leftrightarrow B)$.
The first equality follows from equation (\ref{MFMafmUEI P def expanded Symm AFR})
$\textbf{P}_{A}=\textbf{P}_{B}=\mbox{\boldmath $\delta$}_{1}c_{0,AB}
(\textbf{S}_{A}\cdot\textbf{S}_{B})$.
So, we get $\partial_{t}\textbf{P}= \partial_{t}\textbf{P}_{A}+\partial_{t}\textbf{P}_{B}=2\partial_{t}\textbf{P}_{A}$.
Final result has the following form
$$\partial_{t}\textbf{P}=2\mbox{\boldmath $\delta$}_{1}
g_{0\Pi}g_{(0\zeta_{1})}\times$$
$$\times\biggl( (\mbox{\boldmath $\delta$}_{2,AB}\cdot\textbf{S}_{A})(\textbf{S}_{A}\cdot \textbf{S}_{B})-(\mbox{\boldmath $\delta$}_{2,AB}\cdot\textbf{S}_{B}) (\textbf{S}_{A}\cdot \textbf{S}_{A})$$
$$-(\mbox{\boldmath $\delta$}_{2,AB}\cdot\textbf{S}_{B})(\textbf{S}_{A}\cdot\textbf{S}_{B}) + (\mbox{\boldmath $\delta$}_{2,AB}\cdot\textbf{S}_{A})(\textbf{S}_{B}\cdot \textbf{S}_{B}) \biggr)$$
\begin{equation}\label{MFMemf PEE with D Dz const str G coll}
+2g_{(0\zeta_{1}\Pi)}
\mbox{\boldmath $\delta$}_{1}(\delta_{2,AB}^{\beta}\pi^{\beta\gamma}_{A}S^{\gamma}_{B}-\delta_{2,AB}^{\beta}\pi^{\beta\gamma}_{B}S^{\gamma}_{A}),
\end{equation}
where $\pi^{\beta\gamma}_{A}=\pi^{\beta\gamma}_{A}(\textbf{r},t)$
is the density of nematic tensor:
\begin{equation}\label{MFMemf nematic tensor A def}
\pi_{A}^{\alpha\beta}=\frac{1}{2}\int dR\sum_{i\in A}
\delta(\textbf{r}-\textbf{r}_{i})
\Psi^{\dag}(R,t)(\hat{S}_{i}^{\alpha}\hat{S}_{i}^{\beta}+\hat{S}_{i}^{\beta}\hat{S}_{i}^{\alpha})\Psi(R,t).
\end{equation}
Equation (\ref{MFMemf PEE with D Dz const str G coll})
includes three following interaction constants
$g_{0\Pi}\mbox{\boldmath $\delta$}_{1}=c_{0}\mbox{\boldmath $\delta$}_{1}=\int \mbox{\boldmath $\Pi$}(\xi) d^{3}\xi$,
$g_{(0\zeta_{1})}=\int \zeta_{1}(\xi)d^{3}\xi$,
and
$g_{(0\zeta\Pi)}\mbox{\boldmath $\delta$}_{1}=\int \mbox{\boldmath $\Pi$}(\xi) \zeta(\xi)d^{3}\xi.$

In order to get complete analysis for the polarization evolution equation
we consider the time derivative for the macroscopic approximate form of polarization
(\ref{MFMafmUEI P def expanded Symm AFR}),
where we use the
partial
Landau--Lifshitz--Gilbert equations
(\ref{MFMemf s evolution MAIN TEXT 2 A})
and
(\ref{MFMemf s evolution MAIN TEXT 2 B})
for the time derivatives of the spin densities.
It allows to reproduce the first part of equation (\ref{MFMemf PEE with D Dz const str G coll}),
which is proportional to $g_{0\Pi}g_{(0\zeta_{1})}$.
The last group of terms containing the nematic tensor density does not appear at such derivation.

It is possible to give representation of equation (\ref{MFMemf PEE with D Dz const str G coll}) via the antiferromagnetic vectors
$\textbf{S}_{A}=(\textbf{M}+\textbf{L})/2$ and $\textbf{S}_{B}=(\textbf{M}-\textbf{L})/2$.
To some extend it is possible to introduce the antiferromagnetic nematic tensors as well:
$\pi_{M}^{\alpha\beta}=\pi_{A}^{\alpha\beta}+\pi_{B}^{\alpha\beta}$ ,
and
$\pi_{L}^{\alpha\beta}=\pi_{A}^{\alpha\beta}-\pi_{B}^{\alpha\beta}$.

\subsubsection{Regime of "noncollinear" spins}

In this subsection we apply polarization
(\ref{MFMemf pol A def})
with the electric dipole moment (\ref{MFMemf edm operator NONsimm}).
Using the microscopic many-particle Pauli equation for the description of quantum dynamics in accordance with the quantum hydrodynamic method
\cite{AndreevTrukh PS 24}, \cite{Andreev LP 21 fermions}, \cite{Andreev 2025 Vestn}
we derive corresponding polarization evolution equation
$$\partial_{t}P_{A}^{\mu}=-\frac{1}{3}\varepsilon^{\alpha\beta\gamma}\varepsilon^{\mu\nu\sigma}\varepsilon^{\sigma ab}\varepsilon^{\gamma b\lambda}
\delta_{n,AB}^{\alpha}\times$$
$$\times\Biggl(g_{(0\zeta_{n})}g_{(2\alpha)}[S^{a}_{A}\partial^{\nu}(S_{A}^{\beta}S_{B}^{\lambda})+S_{A}^{\lambda}S_{B}^{\beta}\partial^{\nu}S_{B}^{a}]$$
\begin{equation}\label{MFMemf PEE with D Dz const str G non coll A}
+g_{(2\alpha\zeta_{n})}[\pi_{A}^{\beta a}\partial^{\nu}S_{B}^{\lambda}+S_{A}^{\lambda}\partial^{\nu}\pi_{B}^{a\beta}]
\Biggr),\end{equation}
where we find contribution of three interaction constants
$g_{(0\zeta_{n})}=\int \zeta_{n}(\xi)d^{3}\xi$,
$g_{(2\alpha)}=\int \xi^{2}\alpha(\xi)d^{3}\xi$,
and
$g_{(2\alpha\zeta_{n})}=\int \xi^{2}\alpha(\xi)\zeta_{n}(\xi)d^{3}\xi$.
Constant $g_{(0\zeta_{n})}$ is considered above at the derivation of
the energy density (\ref{MFMemf energy fin 3})
and the Landau--Lifshitz--Gilbert equations (the spin evolution equations)
(\ref{MFMemf s evolution MAIN TEXT 2 A}) and (\ref{MFMemf s evolution MAIN TEXT 2 B}).
Constant $g_{(2\alpha)}$ is described at the analysis of structure of macroscopic polarization (\ref{MFMemf P def expanded noncoll}).
Constant $g_{(2\alpha\zeta_{n})}$ appears in equation (\ref{MFMemf PEE with D Dz const str G non coll A})  for the first time.

Equation (\ref{MFMemf PEE with D Dz const str G non coll A}),
similarly to equation (\ref{MFMemf PEE with D Dz const str G coll}),
contains the contribution of the nematic tensor $\pi_{A}^{\alpha\beta}$ (\ref{MFMemf nematic tensor A def}).
Presence of the nematic tensor $\pi_{A}^{\alpha\beta}$ and novel constant $g_{(2\alpha\zeta_{n})}$
in last term in equation (\ref{MFMemf PEE with D Dz const str G non coll A})
shows that this part of the polarization evolution cannot be derived
via the time derivative of the approximate expression (\ref{MFMemf P def expanded noncoll AFM A})
and application of the Landau--Lifshitz--Gilbert equation with terms
(\ref{MFMemf s evolution MAIN TEXT 2 A})
and
(\ref{MFMemf s evolution MAIN TEXT 2 B}).

We also present the structure of the second partial polarization
\begin{equation}\label{MFMemf PEE with D Dz const str G non coll B}
\partial_{t}\textbf{P}_{B}=-\partial_{t}\textbf{P}_{A}(A\leftrightarrow B),\end{equation}
and full polarization  
\begin{equation}\label{MFMemf PEE with D Dz const str G non coll AB}
\partial_{t}\textbf{P}=\partial_{t}\textbf{P}_{A}+\partial_{t}\textbf{P}_{B}.\end{equation}
Complete expression for the full polarization is not presented explicitly
since there is a rather small simplification of the found expressions at combination of
$\partial_{t}\textbf{P}_{A}$ and $\partial_{t}\textbf{P}_{B}$.
Obviously, equation for $\partial_{t}\textbf{P}$ as well as $\partial_{t}\textbf{P}_{A}$ and $\partial_{t}\textbf{P}_{B}$
can be represented via the antiferromagnetic vectors
$\textbf{S}_{A}=(\textbf{M}+\textbf{L})/2$ and $\textbf{S}_{B}=(\textbf{M}-\textbf{L})/2$.


\subsection{Polarization evolution for the Dzyaloshinskii constant with double vector product}

In this subsection we present another part of the polarization evolution equation
which happens under the action of DMI with the DMC proportional to double vector product of the relative distance and partial ligand shifts
\begin{equation}\label{MFMemf D Dz const str G double vp 5}
\textbf{D}_{ij,(dvp)}=\zeta_{2}(r_{ij})\mbox{\boldmath $\delta$}_{1-AB}(\textbf{r}_{ij}\cdot\mbox{\boldmath $\delta$}_{3-AB}).\end{equation}
We consider two mechanisms of the polarization formation.
Hence, we use the electric dipole moments presented above to define the polarization (\ref{MFMemf pol A def}).
Next, we consider the time derivative of the polarization and replace the time derivatives of the wave functions by the DMI Hamiltonian in accordance with the many-particle Pauli equation.
Below we present corresponding contributions in the polarization evolution equation.

\subsubsection{Regime of "collinear" spins}


In this subsection we consider polarization
(\ref{MFMemf pol A def})
with the electric dipole moment (\ref{MFMemf edm operator simm}).
However, we use the DMI with DMC in form of (\ref{MFMemf D Dz const str G double vp 5}).
It leads to the polarization evolution equation:
$$\partial_{t}P_{A}^{\mu}=\frac{1}{3}\varepsilon^{\alpha\beta\gamma}\varepsilon^{\gamma\nu\sigma}
\delta_{1}^{\mu}\delta_{2,AB}^{\alpha}\delta_{3,AB}^{\lambda}\times$$
$$\times\biggl[c_{0}g_{(2\zeta_{2})} \biggl(S_{A}^{\nu}(\partial^{\lambda}S_{A}^{\beta})S_{B}^{\sigma}+S_{A}^{\sigma}(\partial^{\lambda}S_{B}^{\beta})S_{B}^{\nu}\biggr)$$
\begin{equation}\label{MFMemf PEE with D Dz const str G double vp coll A}
+g_{(2\eta\zeta_{2})}\biggl(S_{A}^{\sigma}\partial^{\lambda}\pi_{B}^{\nu\beta} -\pi_{A}^{\beta\nu}\partial^{\lambda}S_{B}^{\sigma}\biggr)\biggr],
\end{equation}
where
$g_{(2\eta\zeta_{2})}\equiv\int \zeta_{2}(r)\eta(r)r^{2}d^{3}r$
with
$\mbox{\boldmath $\Pi$}=\mbox{\boldmath $\delta$}_{1}\eta$.
other interaction constants
$c_{0}$ and $g_{(2\zeta_{2})}$ are described above.

It is possible to point out that $\partial_{t}\textbf{P}_{A}\parallel \mbox{\boldmath $\delta$}_{1}$.
It is correct for both parts (one constructed of the spin density, another includes the nematic tensor density).
It is also similar to equation (\ref{MFMemf PEE with D Dz const str G coll})
which is also obtained in the regime of polarization formation due to the "collinear" parts of spins,
but for the different form of the DMC.

If we consider the lowest order of the expansion on the small relative distance
we obtain
$$\partial_{t}P_{A}^{\mu}\mid_{0}=\partial_{t}P_{B}^{\mu}\mid_{0}.$$
It also corresponds to $\textbf{P}_{A}=\textbf{P}_{B}$ in the zeroth order of expansion.
So, the application of the Landau--Lifshitz--Gilbert equation repeats this result.
Obviously, the usage of the Landau--Lifshitz--Gilbert equation reproduce the spin density dependent part only,
but it does not give the contribution of the nematic tensor.

\subsubsection{Regime of "noncollinear" spins}


In this subsection we apply general definition of polarization
(\ref{MFMemf pol A def})
with the electric dipole moment (\ref{MFMemf edm operator NONsimm}) corresponding to the regime of "noncollinear" spins.
Our calculations lead to the following form of the polarization evolution equation.
Obtained expression is rather complex,
so it is splitted on several parts.
The first part of the equation for the evolution of the partial polarization is
$$\partial_{t}P_{A(1)}^{\mu}=\frac{1}{2}\frac{1}{3^{3}}\delta_{1,AB}^{\alpha}\delta_{3,AB}^{\delta} g_{(2\alpha)}g_{(2\zeta_{2})}
\varepsilon^{\alpha\beta\gamma}\varepsilon^{\mu\nu\sigma}\varepsilon^{\sigma ab}\varepsilon^{\gamma b\lambda}\times$$
$$\times\biggl[S_{A}^{\beta}\biggl((\partial^{\delta}S_{B}^{\lambda})(\partial^{\nu}S_{A}^{a})
-(\partial^{\nu}S_{B}^{\lambda})(\partial^{\delta}S_{A}^{a}) \biggr)$$
$$+S_{B}^{\lambda}\biggl((\partial^{\delta}S_{A}^{a})(\partial^{\nu}S_{A}^{\beta}) -(\partial^{\nu}S_{A}^{a})(\partial^{\delta}S_{A}^{\beta})\biggr)$$
\begin{equation}\label{MFMemf PEE with D Dz const str G double vp non coll A 1}
+ S_{A}^{a}\biggl((\partial^{\delta}S_{A}^{\beta})(\partial^{\nu}S_{B}^{\lambda}) -(\partial^{\nu}S_{A}^{\beta})(\partial^{\delta}S_{B}^{\lambda})\biggr)
-6S_{A}^{a}\partial^{\nu}(S_{B}^{\lambda}\partial^{\delta}S_{A}^{\beta})\biggr].
\end{equation}
Here, the terms composed of three spin densities include $\textbf{S}_{A}$ twice and $\textbf{S}_{B}$ ones.

The second part of the equation for the evolution of the partial polarization $\textbf{P}_{A}$ differs
from the first part $\partial_{t}P_{A(1)}^{\mu}$ by composition of the spin density dependent terms via partial spin densities.
In the second part we get $\textbf{S}_{A}$ ones and $\textbf{S}_{B}$ twice.
Corresponding expression has the following form:
$$\partial_{t}P_{A(2)}^{\mu}=\frac{1}{6}\frac{1}{3^{2}}\delta_{1,AB}^{\alpha}\delta_{3,AB}^{\delta} g_{(2\alpha)}g_{(2\zeta_{2})}
\varepsilon^{\alpha\beta\gamma}\varepsilon^{\mu\nu\sigma}\varepsilon^{\sigma ab}\varepsilon^{\gamma b\lambda}\times$$
$$\times\biggl[
S_{B}^{a}\biggl(  (\partial^{\delta}S_{B}^{\beta})(\partial^{\nu}S_{A}^{\lambda}) -(\partial^{\nu}S_{B}^{\beta})(\partial^{\delta}S_{A}^{\lambda})\biggr)   $$
$$
+S_{B}^{\beta}  \biggl(  (\partial^{\delta}S_{A}^{\lambda})(\partial^{\nu}S_{B}^{a}) -(\partial^{\nu}S_{A}^{\lambda})(\partial^{\delta}S_{B}^{a})\biggr)
$$
$$
+S_{A}^{\lambda}  \biggl(  (\partial^{\delta}S_{B}^{a})(\partial^{\nu}S_{B}^{\beta}) -(\partial^{\nu}S_{B}^{a})(\partial^{\delta}S_{B}^{\beta})\biggr)
$$
\begin{equation}\label{MFMemf PEE with D Dz const str G double vp non coll A 2}
+6S_{A}^{\lambda} (\partial^{\nu} S_{B}^{a})(\partial^{\delta}S_{B}^{\beta}) \biggr].
\end{equation}
First three group of terms in $\partial_{t}P_{A(2)}^{\mu}$ (\ref{MFMemf PEE with D Dz const str G double vp non coll A 2})
are similar to the first three group of terms in $\partial_{t}P_{A(1)}^{\mu}$ (\ref{MFMemf PEE with D Dz const str G double vp non coll A 1})
at exchange of subsystem indexes $A\leftrightarrow B$.
But the last terms in equations (\ref{MFMemf PEE with D Dz const str G double vp non coll A 1})  and
(\ref{MFMemf PEE with D Dz const str G double vp non coll A 2}) have different structures.

The third part of $\partial_{t}\textbf{P}_{A}$ contains the contribution of the nematic tensor density:
$$\partial_{t}P_{A(3)}^{\mu}=\frac{1}{3}\delta_{1,AB}^{\alpha}\delta_{3,AB}^{\delta}
\varepsilon^{\alpha\beta\gamma}\varepsilon^{\mu\nu\sigma}\varepsilon^{\sigma ab}\varepsilon^{\gamma b\lambda}\times$$
$$\times\biggl( g_{(2\zeta_{2}\alpha)}(\pi_{A}^{a\beta} S_{B}^{\lambda} + S_{A}^{\lambda} \pi_{B}^{a\beta})
+\frac{1}{10}g_{(4\zeta_{2}\alpha)}
(\pi_{A}^{a\beta} \delta^{\delta\nu }\triangle S_{B}^{\lambda}$$
\begin{equation}\label{MFMemf PEE with D Dz const str G double vp non coll A 3}
+2\pi_{A}^{a\beta}\partial^{\delta}\partial^{\nu} S_{B}^{\lambda}
+ S_{A}^{\lambda} \delta^{\delta\nu }\triangle \pi_{B}^{a\beta}+2 S_{A}^{\lambda}\partial^{\delta}\partial^{\nu} \pi_{B}^{a\beta}) \biggr),
\end{equation}
where
$g_{(4\zeta_{2}\alpha)}=\int \zeta_{2}(r)\alpha(r)r^4 d^3r$.
It also contains the contribution of the zeroth order expansion
(it is proportional to $g_{(2\zeta_{2}\alpha)}$).
This term gives major contribution in $\partial_{t}\textbf{P}_{A}$ since all other terms
(including $\partial_{t}\textbf{P}_{A(1)}$ and $\partial_{t}\textbf{P}_{A(2)}$) are found in the second order expansion.

An analog of $\partial_{t}P_{A(2)}^{\mu}$ on the microscopic structure in $\partial_{t}P_{B}^{\mu}$ appears to be $\partial_{t}P_{B(1)}^{\mu}$.
Its macroscopic structure can be presented in short inexplicit form:
\begin{equation}\label{MFMemf PEE with D Dz const str G double vp non coll B 1}
\partial_{t}P_{B(1)}^{\mu}=\partial_{t}P_{A(2)}^{\mu}(A\leftrightarrow B).
\end{equation}

The second partial contribution in $\partial_{t}P_{B}^{\mu}$ is presented in explicit form:
$$\partial_{t}P_{B(2)}^{\mu}=\frac{1}{6}\frac{1}{3^{2}}\delta_{1,AB}^{\alpha}\delta_{3,AB}^{\delta} g_{(2\alpha)}g_{(2\zeta_{2})}
\varepsilon^{\alpha\beta\gamma}\varepsilon^{\mu\nu\sigma}\varepsilon^{\sigma ab}\varepsilon^{\gamma b\lambda}\times$$
$$\times\biggl[
S_{B}^{a}\biggl(  (\partial^{\delta}S_{B}^{\beta})(\partial^{\nu}S_{A}^{\lambda}) -(\partial^{\nu}S_{B}^{\beta})(\partial^{\delta}S_{A}^{\lambda})\biggr)$$
$$+S_{B}^{\beta}  \biggl(  (\partial^{\delta}S_{A}^{\lambda})(\partial^{\nu}S_{B}^{a}) -(\partial^{\nu}S_{A}^{\lambda})(\partial^{\delta}S_{B}^{a})\biggr)$$
\begin{equation}\label{MFMemf PEE with D Dz const str G double vp non coll B 2}
+S_{A}^{\lambda}  \biggl(  (\partial^{\delta}S_{B}^{a})(\partial^{\nu}S_{B}^{\beta}) -(\partial^{\nu}S_{B}^{a})(\partial^{\delta}S_{B}^{\beta})\biggr)
-6S_{B}^{a} \partial^{\nu} (S_{A}^{\lambda}\partial^{\delta}S_{B}^{\beta}) \biggr].
\end{equation}
Comparison with equation (\ref{MFMemf PEE with D Dz const str G double vp non coll A 1}) shows
that the relation similar to (\ref{MFMemf PEE with D Dz const str G double vp non coll B 1})
is correct in this case as well
$\partial_{t}P_{B(2)}^{\mu}=\partial_{t}P_{A(1)}^{\mu}(A\leftrightarrow B)$.

The last part of the polarization evolution has the following structure
\begin{equation}\label{MFMemf PEE with D Dz const str G double vp non coll b 3}
\partial_{t}P_{B(3)}^{\mu}=\partial_{t}P_{A(3)}^{\mu}(A\leftrightarrow B).
\end{equation}
Moreover, the first terms, appearing in the zeroth order of the expansion,
in $\partial_{t}P_{A(3)}^{\mu}$ and $\partial_{t}P_{B(3)}^{\mu}$ are equal to each other
$\partial_{t}P_{B(3)}^{\mu}=\partial_{t}P_{A(3)}^{\mu}$.

\section{Dzyaloshinskii-Moriya interaction contribution in the Euler equation for the momentum density}

In this section we consider the contributions of the Dzyaloshinskii-Moriya interaction
in the macroscopic Euler equation for the velocity field,
corresponding to the different parts of the Dzyaloshinskii constant.

The Euler equation is the necessary part of the model if the acoustic waves are considered along with the spin-waves.
Moreover, the balance between the electric dipole-dipole interaction and the spin-orbit interaction in the Euler equation leads to the spin-current model of polarization.
Consequently, the calculation of the Dzyaloshinskii-Moriya interaction in the Euler equation is required for the complete analysis of the magnetic properties of the medium.

First, we need to present the definition of the momentum density and the concentration (the number density)
in terms of material-field representation of many-particle quantum mechanics
\cite{Andreev PoF 21}, \cite{Andreev LP 21 fermions}, \cite{Andreev 2025 Vestn}:
\begin{equation}\label{MFMemf concentration A def}
n_{A}=\int dR\sum_{i\in A}
\delta(\textbf{r}-\textbf{r}_{i})
\Psi^{\dag}(R,t)\Psi(R,t)
\end{equation}
for the partial concentration of ions with one of opposite spin projections,
and
\begin{equation}\label{MFMemf momentum density A def}
m\textbf{j}_{A}=\frac{1}{2}\int dR\sum_{i\in A}
\delta(\textbf{r}-\textbf{r}_{i})
[\Psi^{\dag}(R,t)\hat{\textbf{p}}_{i}\Psi(R,t)+h.c.]
\end{equation}
for the partial momentum density,
where $\hat{\textbf{p}}_{i}=-\imath\hbar\nabla_{i}$ is the momentum operator for i-th particle,
and
h.c. is the Hermitian conjugation.
Equations (\ref{MFMemf concentration A def}) and (\ref{MFMemf momentum density A def})
contain notations similar to the spin density definition (\ref{MFMemf spin A def}).

\subsection{The Keffer form of the Dzyaloshinskii constant}

Here, we use the DMC
$\textbf{D}_{ij}=\gamma(r_{ij})\textbf{r}_{ij}
+\beta(r_{ij})[\textbf{r}_{ij}\times\mbox{\boldmath $\delta$}_{1}]$
in order to derive the force density related to DMI.
We can represent this DMC as the single term
$D_{ij}^{\alpha}\equiv\kappa^{\alpha\beta}r_{ij}^{\beta}$,
where $\kappa^{\alpha\beta}(r)=\gamma(r) \delta^{\alpha\beta}+\beta(r)\varepsilon^{\alpha\beta\gamma}\delta_{1}^{\gamma}$.

In the lowest (zeroth) order of expansion on the small relative distance
we find
$$m\partial_{t}j_{lowest,A}^{\mu}=
-\varepsilon^{\alpha\beta\gamma}S_{A}^{\beta}S_{B}^{\gamma}\times$$
\begin{equation}\label{MFMemf v evolution 1 low}
\times\int\biggl(\kappa^{\alpha\mu}+r^{\delta}r^{\mu}\frac{1}{r}\frac{\partial \kappa^{\alpha\delta}}{\partial r}\biggr)d^{3}r
,\end{equation}
where indexes of $\kappa^{\alpha\mu}$ does no related to vector $r^{\mu}$,
with
$\int\biggl(\kappa^{\alpha\mu}+r^{\delta}r^{\mu}\frac{1}{r}\frac{\partial \kappa^{\alpha\delta}}{\partial r}\biggr)d^{3}r$
$=\int\biggl(\kappa^{\alpha\mu}+\frac{1}{3}r\frac{\partial \kappa^{\alpha\mu}}{\partial r}\biggr)d^{3}r$.
The last term can be integrated by parts
$\int\frac{1}{3}r\frac{\partial \kappa^{\alpha\mu}}{\partial r}d^{3}r=-\int\kappa^{\alpha\mu}d^{3}r$.
So, this integral and the force field (\ref{MFMemf v evolution 1 low}) are equal to zero.

For the derivation of nonzero force field we need to consider the next order of expansion
(it is the second order)
$$m\partial_{t}j_{A}^{\mu}= -\frac{1}{2}\varepsilon^{\alpha\beta\gamma} (S_{A}^{\beta}\partial^{a}\partial^{b}S_{B}^{\gamma})\times$$
\begin{equation}\label{MFMemf v evolution 1}
\times
\int r^{a}r^{b}\biggl(\kappa^{\alpha\mu}+r^{\delta}r^{\mu}\frac{1}{r}\frac{\partial \kappa^{\alpha\delta}}{\partial r}\biggr)d^{3}r
.\end{equation}
Complete understanding of the vector structure of the found equation requires the simplification of the integral part of this equation.
It can be simplified in the following way:
$\int r^{a}r^{b}\kappa^{\alpha\mu}d^{3}r$
$=\frac{1}{3}\delta^{ab}\int r^{2}\kappa^{\alpha\mu}d^{3}r$
for the first part of the integral,
and
$\int r^{a}r^{b}r^{\delta}r^{\mu}\frac{1}{r}\frac{\partial \kappa^{\alpha\delta}}{\partial r}d^{3}r$
$=\frac{1}{15}(\delta^{ab}\delta^{\mu\delta}+\delta^{a\mu}\delta^{b\delta}+\delta^{a\delta}\delta^{b\mu})
\int r^{3}\frac{\partial \kappa^{\alpha\delta}}{\partial r}d^{3}r$
$=-\frac{1}{3}(\delta^{ab}\delta^{\mu\delta}+\delta^{a\mu}\delta^{b\delta}+\delta^{a\delta}\delta^{b\mu})
\int r^{2}\kappa^{\alpha\delta}d^{3}r$
for the second part.
We substitute the combined contribution of these integrals in equation (\ref{MFMemf v evolution 1}):
$$m\partial_{t}j_{A}^{\mu}= -\frac{1}{2}\varepsilon^{\alpha\beta\gamma} (S_{A}^{\beta}\partial^{a}\partial^{b}S_{B}^{\gamma})\times$$
$$\times
(-1/3)(\delta^{a\mu}\delta^{b\delta}+\delta^{a\delta}\delta^{b\mu})\int r^{2}\kappa^{\alpha\mu}d^{3}r=$$
\begin{equation}\label{MFMemf v evolution 1 mid}
=\frac{1}{3}\varepsilon^{\alpha\beta\gamma} (S_{A}^{\beta}\partial^{\mu}\partial^{\delta}S_{B}^{\gamma})\int r^{2}\kappa^{\alpha\mu}d^{3}r
.\end{equation}

To finalize our analysis, we need to include the structure $\kappa^{\alpha\mu}$:
$$m\partial_{t}j_{A}^{\mu}=\frac{1}{3} \varepsilon^{\alpha\beta\gamma}\times$$
\begin{equation}\label{MFMemf v evolution 1 fin}
\times\biggl(g_{(2\gamma)}(S_{A}^{\beta}\partial^{\mu}\partial^{\alpha}S_{B}^{\gamma})
+g_{(2\beta)}(S_{A}^{\beta}[\mbox{\boldmath $\delta$}_{1}\times\nabla]^{\alpha}\partial^{\mu}S_{B}^{\gamma})
\biggr)
.\end{equation}
This result is a generalization of equations 8 and 11 in Ref. \cite{AndreevTrukh JETP 24}
(which corresponds to the limit regime of the second term in (\ref{MFMemf v evolution 1 fin}), for the ferromagnetic materials).

\subsection{Generalized Keffer form of the Dzyaloshinskii constant for AFM}

It is demonstrated above, that the following DMC
\begin{equation}\label{MFMemf D Dz const str G for vEE}
\textbf{D}_{ij}=\zeta_{1}(r_{ij})\mbox{\boldmath $\delta$}_{2,ij-AB}
\end{equation}
corresponds to the energy density (\ref{MFMemf energy fin 3}) suggested
by
Dzyaloshinskii.
Here we derive the force field corresponding to this part of DMI:
\begin{equation}\label{MFMemf v evolution 2}
m\partial_{t}j_{A}^{\mu}=-\varepsilon^{\alpha\beta\gamma}g_{(0\zeta_{1})}
\delta_{2,AB}^{\alpha}S_{A}^{\beta}\partial^{\mu}S_{B}^{\gamma}
.\end{equation}
In contrast to the previous subsection, we have different dependence on $r^{\mu}_{ij}$.
Hence, the force field appears in the first order expansion.

\subsection{Novel parts of the Dzyaloshinskii constant for AFM}

The fourth and last part of the force field corresponds to the DMC of form of
$\textbf{D}_{ij,(dvp)}=\zeta_{2}(r_{ij})\mbox{\boldmath $\delta$}_{1-AB}(\textbf{r}_{ij}\cdot\mbox{\boldmath $\delta$}_{3-AB})$.
Its structure is similar to one considered in Sec. 7.A.
So general conclusions are similar as well.
Therefore, we present the force field appearing in this regime:
$$m\partial_{t}j_{A}^{\mu}=  \frac{1}{3} \varepsilon^{\alpha\beta\gamma}\times $$
\begin{equation}\label{MFMemf v evolution 3}
\times g_{(2\zeta_{2})} \delta_{1,AB}^{\alpha}
S_{A}^{\beta}(\mbox{\boldmath $\delta$}_{3-AB}\cdot\nabla)\partial^{\mu}S_{B}^{\gamma}
.\end{equation}

Equations (\ref{MFMemf v evolution 1 fin}), (\ref{MFMemf v evolution 2}), and (\ref{MFMemf v evolution 3})
give the full contribution of the DMI in the Euler equation for one of subspecies $A$.
The force field for the second subspecies and the representation of the force fields via $\textbf{L}$ and $\textbf{M}$ can be given as well.

\section{On a possibility of the Keffer-like form of the exchange integral in symmetric Heisenberg Hamiltonian}

The Heisenberg Hamiltonian or the symmetric exchange interaction is
the symmetric part of the general exchange tensor integral existing in the superexchange regime,
where some form of the exchange interaction of magnetic ions happens via nonmagnetic ion located between magnetic ions.
Another part of the exchange tensor integral (describing superexchange interaction) is its antisymmetric part,
which is the Dzyaloshinskii-Moriya interaction.
The shift of the nonmagnetic ion (ligand) from the middle point between magnetic ions gives contribution in the Dzyaloshinskii constant.
So, we can expect that the ligand shift gives some contribution in the symmetric exchange interaction.

We can consider some generalizations of the "exchange integral" related to some symmetric contributions of the ligand ion shift.
First, we present corresponding Heisenberg Hamiltonian
\begin{equation}\label{MFMemf Ham HHI}
\hat{H}=-\frac{1}{2}\sum_{i,j,j\neq i}U_{ij}(\hat{\textbf{S}}_{i}\cdot \hat{\textbf{S}}_{j}),
\end{equation}
where function $U_{ij}=U(r_{ij})$ is the exchange integral.

$$U_{ij}=U_{0,ij}+U_{1,ij}(\textbf{r}_{ij}\cdot\mbox{\boldmath $\delta$}_{n,ij-AB}) $$
$$+U_{1,ij}'(\mbox{\boldmath $\delta$}_{1,ij}\cdot[\textbf{r}_{ij}\times\mbox{\boldmath $\delta$}_{2,ij-AB})]) $$
\begin{equation}\label{MFMemf U structure in Ham HHI}
+U_{2,ij}(\textbf{r}_{ij}\cdot\mbox{\boldmath $\delta$})^{2},
\end{equation}
where it is possible to include terms like
$U_{3,ij}(\mbox{\boldmath $\delta$}\cdot\mbox{\boldmath $\delta$})$
and
$U_{4,ij}(\mbox{\boldmath $\delta$}\cdot\mbox{\boldmath $\delta$})(\textbf{r}_{ij}\cdot\textbf{r}_{ij})$,
but they give renormalization of $U_{0,ij}$.
In equation (\ref{MFMemf U structure in Ham HHI})
we present the following terms:
$U_{0,ij}$ is the usual exchange integral,
it presents the term containing no contribution of the ligand shift.
Main interest is focused on the second term
$U_{1,ij}(\textbf{r}_{ij}\cdot\mbox{\boldmath $\delta$}_{n,ij-AB})$
and the third term
$U_{1,ij}'(\mbox{\boldmath $\delta$}_{n,ij}\cdot[\textbf{r}_{ij}\times\mbox{\boldmath $\delta$}_{n',ij-AB}])$,
which exists for the two-component magnetic structures like the antiferromagnetic materials.
These terms give qualitatively new structure in the spin torques.
However, full analysis of this contribution will be presented in another paper.
Here, it is mentioned in comparison with the contribution of the ligand shifts in the DMC.
The last term leads to the renormalization of the coefficient in front of the second derivative in the Landau--Lifshitz--Gilbert equation.
They also give some anisotropy of this coefficient,
which appears in addition (or instead) of the anisotropy usually described via $X,Y,Z$-model.

\section{Conclusion}

Different shifts of the ligand ions from the center of mass of the pair of neighboring magnetic ions have been considered
for the ferromagnetic and antiferromagnetic materials.
Corresponding partial shifts have been introduced.
These shifts have been included in the Dzyaloshinskii vector constant in the microscopic Dzyaloshinskii-Moriya interaction
and in the exchange integral in the symmetric exchange Heisenberg Hamiltonian.
Corresponding generalizations of the Keffer form of the Dzyaloshinskii constant have been suggested
in order to obtain novel spin torque in
the Landau--Lifshitz--Gilbert equation,
additional terms in the energy density,
corresponding force density in the momentum balance Euler equation.
It basically has provided
corresponding generalizations of the Dzyaloshinskii-Moriya interaction.

Corresponding parts of the polarization evolution equation associated with the Dzyaloshinskii-Moriya interaction have been derived.
Particularly, let us to point out that
the polarization evolution equation has been consider for the Dzyaloshinskii-Moriya interaction associated with its macroscopic form of energy proportional to the vector product of the partial spin densities of two sublattices in AFM materials with no derivatives.

Presented theoretical background provides the possibility of the systematic analysis of the Dzyaloshinskii-Moriya interaction
and its consequences related to the ligands similar to the Keffer form of the Dzyaloshinskii constant.
It also opens a possibility for more general analysis of many macroscopic phenomena
like spin waves, spin solitons, skyrmions etc in magnetic structures.

\section{DATA AVAILABILITY}

Data sharing is not applicable to this article as no new data were
created or analyzed in this study, which is a purely theoretical one.

\section{Acknowledgements}

The work is supported by the Russian Science Foundation under the
grant
No. 25-22-00064.


\end{document}